\newcommand{\be}{\begin{equation}}
\newcommand{\ee}{\end{equation}}
\newcommand{\bfi}{\begin{figure}}
\newcommand{\efi}{\end{figure}}
\newcommand{\bea}{\begin{eqnarray}}
\newcommand{\eea}{\end{eqnarray}}
\newcommand{\beann}{\begin{eqnarray*}}
\newcommand{\eeann}{\end{eqnarray*}}
\newcommand{\beasn}{\begin{sneqnarray}}
\newcommand{\eeasn}{\end{sneqnarray}}
\newcommand{\ba}{\begin{array}}
\newcommand{\ea}{\end{array}}
\newcommand{\nn}{\nonumber}
\newcommand{\Appendix}[1]%
    {\renewcommand{\thesection}{Appendix~\Alph{section}:}%
     \section{#1}}%

\catcode`@=11
\long\def\@makecaption#1#2{
   \vskip 10pt
   \setbox\@tempboxa\hbox{{\small\bf #1.} \ {\small #2}}
   \ifdim \wd\@tempboxa >\hsize       
   {\small\bf #1.} \ {\small #2}\par  
   \else                              
        \hbox to\hsize{\hfil\box\@tempboxa\hfil}
   \fi}
\catcode`@=12

\catcode`@=11
\def\secteqno{\@addtoreset{equation}{section}%
\def\theequation{\thesection.\arabic{equation}}}
\def\endsecteqno{\def\theequation{\@ifundefined{chapter}%
{\arabic{equation}}{\thechapter.\arabic{equation}}}}
\newcounter{subequation}
\def\thesubequation{\alph{subequation}}
\def\sneqnarray{\stepcounter{equation}\let\@currentlabel=\theequation
\setcounter{subequation}{1}
\def\@eqnnum{{\rm (\theequation\thesubequation)}}
\global\@eqcnt\z@\tabskip\@centering\let\\=\@eqncr\let\@@eqncr=\@@sneqncr
$$\halign to \displaywidth\bgroup\@eqnsel\hskip\@centering
 $\displaystyle\tabskip\z@{##}$&\global\@eqcnt\@ne
 \hskip 2\arraycolsep \hfil${##}$\hfil
 &\global\@eqcnt\tw@ \hskip 2\arraycolsep
$\displaystyle\tabskip\z@{##}$\hfil
  \tabskip\@centering&\llap{##}\tabskip\z@\cr}
\def\endsneqnarray{\@@sneqncr\egroup $$\global\@ignoretrue}
\def\@@sneqncr{\let\@tempa\relax
   \ifcase\@eqcnt \def\@tempa{& & &}\or \def\@tempa{& &}
   \else \def\@tempa{&}\fi
     \@tempa \if@eqnsw\@eqnnum\stepcounter{subequation}\fi
     \global\@eqnswtrue\global\@eqcnt\z@\cr}
\def\nobiblabels{\def\@lbibitem[##1]##2{\@bibitem{##2}}}
\catcode`@=12

\def\a{\alpha}  \def\b{\beta}  
\def\d{\delta}  \def\e{\epsilon}
   
  \def\L{\Lambda} \def\m{\mu} 
  \def\p{\pi}  \def\r{\rho}
\def\s{\sigma} \def\t{\tau}

\def\pa{\partial}  

\documentclass[12pt]{article}
\usepackage[german,english]{babel}
\usepackage{amssymb}
\usepackage{amsmath}   
\usepackage{braket}                                                      
\usepackage{slashed}
\usepackage{multirow}
\usepackage{epsfig,boxedminipage}

\secteqno
\renewcommand{\thesection}{\arabic{section}.}

\renewcommand{\theequation}{\thesection  \arabic{equation}}

\begin{document}


\title{{\bf Effective Field Theory with Dibaryon Fields: Nucleon-Nucleon amplitudes at N$^2$LO}} 
\author{{\Large {\sl Joan Soto}}  {\sl and} {\Large {\sl Jaume Tarr\'us}}\\
        \small{\it{Departament d'Estructura i Constituents de la Mat\`eria 
                   and Institut de Ci\`encies del Cosmos}}\\
        \small{\it{Universitat de Barcelona}}\\
        \small{\it{Diagonal, 647, E-08028 Barcelona, Catalonia, Spain.}}\\  \\
        {\it e-mails:} \small{tarrus@ecm.ub.es, joan.soto@ub.edu} }
\date{\today}

\maketitle

\thispagestyle{empty}

\begin{abstract}
We calculate the nucleon-nucleon scattering amplitudes in the $^1S_0$ and $^3S_1$-$^3D_1$ channels at next-to-next to leading order starting from a recently proposed non-relativistic chiral effective theory, which includes dibaryon fields as fundamental degrees of freedom. We restrict ourselves to center of mass energies ($E$) smaller than the pion mass ($m_\pi$), and further divide the calculation into two relative momentum ($p$) regions, a high energy one $p\sim m_\pi  \gg \delta m_i$, $\delta m_i $ being the dibaryon residual masses, and low energy one $p\lesssim  \delta m_i$. We first match to a lower energy effective theory in which we calculate the amplitudes in the high energy region. We further match this effective theory to the so called pionless effective theory in the low energy region, and carry out the calculations in the latter. Dimensional regularization and minimal subtraction scheme are used throughout. For $^1S_0$ channel a good description of the phase shift data is obtained for $E\lesssim 50 MeV.$ For the $^3S_1-^3D_1$ channel, the $^3S_1$ phase shift data is only well described up to $E \lesssim 20 MeV.$

\end{abstract}

PACS: 14.20.Pt, 13.75.Cs, 21.30.Fe, 21.45.Bc, 03.65.Nk .

\vfill
\vbox{
\hfill{}\null\par
\hfill{UB-ECM-PF 09/11}\null\par}

\newpage


\section{Introduction}
\indent

Since the original suggestion by Weinberg \cite{Weinberg} that the nuclear forces could be understood within the framework of effective field theories (EFT) there has been an enormous development of the subject (see 
\cite{Bernard:1995dp,Bedaque:2002mn,Epelbaum:2005pn,Hammer:2006qj,Machleidt:2007ms,Epelbaum:2008ga}
for reviews). A key ingredient of the EFT formalism is that the cut-off dependence which is introduced in order to smooth out ultraviolet (UV) singularities can be absorbed by suitable counterterms, and hence any dependence on physical scales much higher than the ones of the problem at hand can be encoded in a few (unknown) constants. In order to achieve this in a systematic manner counting rules are also necessary.

In a recent paper \cite{Soto:2007pg} we proposed a chiral non-relativistic EFT which included two dibaryon fields as fundamental degrees of freedom. This EFT, which will be simply referred as NNEFT in this paper, is renormalizable and has simple counting rules when dimensional regularization (DR) and minimal subtraction (MS) scheme are used. The nucleon-nucleon scattering amplitudes in the $^1S_0$ and $^3S_1$ channels were calculated at NLO and a good description of data achieved in the $0-50 MeV$ energy range. We carry out here the calculation at N$^2$LO in order to see if the good description of data persists and check the convergence of the EFT. This is mandatory in view of the fact that the so called KSW approach \cite{KSW} also produced a good description of data at NLO, but turned out to have a bad convergence in the $^3S_1$ channel at N$^2$LO \cite{Fleming:1999ee,Fleming:1999bs}.   
We will restrict ourselves to an energy range $E$ such that $E\ll m_\pi$, the pion mass, and $p=\sqrt{m_N E}\sim m_\pi$, $m_N$ being the nucleon mass. Pion fields can then be integrated out leading to an EFT, which was already described in \cite{Soto:2007pg}, which we will call potential NNEFT (pNNEFT). For $p\sim m_\pi$ this EFT will already be suitable to carry out the calculations of the amplitudes. For $p\ll m_\pi$ however it will be convenient to integrate out nucleon fields with $p\sim m_\pi$ and use the so called pionless NNEFT ($\slashed{\p}$NNEFT) \cite{Kaplan:1996xu,vanKolck:1998bw,Beane:2000fi}. All matching calculations will be done expanding the low energy or momentum scales in the integrals and using DR to regulate any possible IR divergence. Local field redefinitions which respect the counting will be used to get ride of redundant operators, rather than using the on-shell condition.

We will organize the paper as follows. In sections 2 and 3 we introduce the NNEFT, 
and the pNNEFT 
Lagrangians respectively. In section 4 we match NNEFT to pNNEFT. In section 5 we calculate the nucleon-nucleon scattering amplitudes in pNNEFT. In section 6 we match pNNEFT to $\slashed{\p}$NNEFT, and calculate the nucleon-nucleon scattering amplitudes in the latter. Sections 7 and 8 are devoted to the comparison of our results with data in the $^1S_0$ and $^3S_1$-$^3D_1$ channels respectively. We close with a discussion and conclusions in section 9.


\section{The nucleon-nucleon chiral effective theory with \\ dibaryon fields} 
\indent

Our starting point is the effective theory for the $N_B$=2 sector of QCD for non-relativistic energies much smaller than $\Lambda_\chi$ 
recently proposed in \cite{Soto:2007pg}. The distinct feature of this EFT is that in addition to the usual degrees of freedom for a NNEFT theory, namely nucleons and pions, two dibaryon fields, an isovector ($D^a_s$) with quantum numbers $^1 S_0$ and an isoscalar ($\vec{D}_v$) with quantum numbers $^3 S_1$ are also included. Since $m_N \sim \Lambda_\chi$, a non-relativistic formulation of the nucleon fields is convenient \cite{JM}. Chiral symmetry, and its breaking due to the quark masses in QCD, constrain the possible interactions of the nucleons and dibaryon fields with the pions. The $N_B=0$ sector is given by the chiral Lagrangian, which will only be needed at LO,
\be
\mathcal{L}_{\pi}=\frac{f_{\pi}^2}{8}\left\lbrace Tr(\pa_{\mu}U^{\dag}\pa^{\mu}U)+2B_0 Tr(\mathcal{M}U^{\dag}+U\mathcal{M}^{\dag})\right\rbrace, \quad U=e^{2i\frac{\pi^a\t^a}{f_{\pi}}}
\ee
$\mathcal{M}$ is the quark mass matrix, which we will take in the isospin limit, namely the average of the up and down quark masses $m_q$ times the identity matrix. $B_0$ is defined by $m^2_{\p}=2B_0m_q$. The $N_B=1$ sector contains the pion-nucleon interactions
, and will be needed at NLO \cite{Weinberg},
\be
\mathcal{L}_{\pi N}=\mathcal{L}_{\pi N}^{(1)}+\mathcal{L}_{\pi N}^{(2)}+\cdots.
\ee

At LO we have,
\be 
\begin{split}
\mathcal{L}_{\pi N}^{(1)}=& N^{\dag}\Bigl(iD_0-g_A(\vec{u}\cdot\frac{\vec{\s}}{2})+\frac{\vec{D}^2}{2m_N}\Bigr)N
\end{split}
\ee
where $u^2=U$, $u_{\m}=i\left\lbrace u^{\dag},\pa_{\m}u\right\rbrace $, $D_{\m}=(\pa_{\m}+\frac{1}{2}[u^{\dag},\pa_{\m}u])$, $\pi^a$ is the pion field, $\t^a$ the isospin Pauli matrices, $g_A\sim 1.25$ is the axial vector coupling constant of the nucleon, and
$f_{\pi}\sim 132 MeV$ is the pion decay constant.
This is the leading order Lagrangian for 
the pion-nucleon interactions ($\mathcal{O}(p)$)
for $E\sim p^2/2m_N \sim m_\pi$. The NLO Lagrangian in this sector reads \cite{Bernard:1995dp},

\be
\begin{split}
\mathcal{L}_{\pi N}^{(2)}= N^{\dag}\biggl\{&\frac{\vec{D}^4}{8m_N^3} 
-\frac{ig_A}{4m_N} \{\vec{\s}\cdot \vec{D},u_0\}+c_1 Tr(\chi_+) +\\
&+ c_2u_0^2 - c_3 \vec{u} \cdot \vec{u}+ i\frac{c_4}{2}\e^{ijk}\s^k u_i u_j +c_5\chi_+
\biggr\}N 
\end{split}
\ee
with $ \chi_+= 2B_0(u^\dagger \mathcal{M} u^\dagger + u \mathcal{M}^\dagger u)$. 
All parameters, $g_A, m_N, c_i, \kappa_{s,v}$ are understood as the ones in the chiral limit.

The $N_B=2$ sector consist of terms with (local) two nucleon interactions, dibaryons and dibaryon-nucleon interactions. The terms with two nucleon interactions can be removed by local field redefinitions \cite{Bedaque:1997qi,Bedaque:1999vb,Beane:2000fi} and will not be further considered.  
The terms with dibaryon fields and no nucleons in the rest frame of the dibaryons read
\be
\mathcal{L}_{D}=\mathcal{L}_{\mathcal{O}(p)}+\mathcal{L}_{\mathcal{O}(p^2)}
\ee
where $\mathcal{L}_{\mathcal{O}(p)}$ is the $\mathcal{O}(p)$ Lagrangian\footnote{The last term starts contributing to the nucleon-nucleon amplitudes at N$^2$LO and, hence, was not displayed in \cite{Soto:2007pg}} ,
\be
\mathcal{L}_{\mathcal{O}(p)}={1\over 2} Tr\left[D_{s}^{\dag}\Bigl(-id_0+\delta_{m_s}'\Bigr)D_s\right]+
\vec{D}_v^{\dag}\Bigl(-i\pa_0+\delta_{m_v}'\Bigr)\vec{D}_v
+ic_{sv}\left(\vec{D}_v^{\dag} Tr \left( {\vec u} D_s\right) - h.c.\right)
\label{dbLO}
\ee
where $D_s=D^a_s\t_a$ and $\delta_{m_i}'$, $i=s,v$ are the dibaryon residual masses, which must be much smaller than $\Lambda_\chi$, otherwise the dibaryon should have been integrated out as the remaining resonances have. The negative signs of the time derivatives are chosen this way in order to eventually reproduce the signs of the effective range parameters. As discussed in \cite{Soto:2007pg}, they do not imply any violation of unitarity.
 
The covariant derivative for the scalar (isovector) dibaryon field is defined as
$d_0D_s=\pa_0D_s+\frac{1}{2}[[u,\pa_0u],D_s]$.
$\mathcal{L}_{\mathcal{O}(p^2)}$ is the $\mathcal{O}(p^2)$ Lagrangian,
\be
\begin{split}
\mathcal{L}_{\mathcal{O}(p^2)}=
&s_1Tr[D_s(u\mathcal{M}^{\dag}u+u^{\dag}\mathcal{M}u^{\dag})D^{\dag}_s]+s_2Tr[D^{\dag}_s(u\mathcal{M}^{\dag}u+u^{\dag}\mathcal{M}u^{\dag})D_s]+\\
&+v_1\vec{D}^{\dag}_v\cdot\vec{D}_vTr[u^{\dag}\mathcal{M}u^{\dag}+u\mathcal{M}^{\dag}u] +\cdots
\end{split}
\label{ordrep2}
\ee
$s_i$, $i=1,2$,
and $v_1$, 
are low energy constants (LEC). We have only displayed here the terms which will eventually contribute in our calculations. The complete list of operators is given in the Appendix B.

The dibaryon-nucleon interactions will also be needed at NLO,
\be
\mathcal{L}_{DN}=\mathcal{L}_{DN}^{(1)}+\mathcal{L}_{DN}^{(2)}+\cdots
\label{dn0}
\ee
At LO they read
\be
\begin{split}
\mathcal{L}_{DN}^{(1)}=&\frac{A_s}{\sqrt{2}}(N^{\dag}\s^2\t^a\t^2N^*)D_{s,a}+
\frac{A_s}{\sqrt{2}}(N^{\top}\s^2\t^2\t^aN)D^{\dag}_{s,a}+\\
&+\frac{A_v}{\sqrt{2}}(N^{\dag}\t^2\vec{\s}\s^2N^*)\cdot\vec{D}_v+\frac{A_v}{\sqrt{2}}(N^{\top}\t^2\s^2\vec{\s} N)\cdot\vec{D}_v^{\dag}
 \end{split}
\label{dn}
\ee
$A_i\sim \Lambda_\chi^{-1/2}$, $i=s,v$, and at NLO
\be
\begin{split}
\mathcal{L}_{DN}^{(2)}=&\frac{B_s}{\sqrt{2}}(N^{\dag}\s^2\t^a\t^2{\vec D}^2N^*)D_{s,a}+\frac{B_s}{\sqrt{2}}(N^{\top}\s^2\t^2\t^a{\vec D}^2N)D^{\dag}_{s,a}+
\\
&+\frac{B_v}{\sqrt{2}}(N^{\dag}\t^2\vec{\s}\s^2{\vec D}^2N^*)\cdot\vec{D}_v+\frac{B_v}{\sqrt{2}}(N^{\top}\t^2\s^2\vec{\s}{\vec D}^2 N)\cdot\vec{D}_v^{\dag}
\\
&+\frac{B'_v}{\sqrt{2}} (D_i N^{\dag}\t^2\s^i\s^2D_j N^*)D^j_v + \frac{B'_v}{\sqrt{2}}(D_i N^{\top}\t^2 \s^2 \s^iD_j N) D_v^{j\dag}
\end{split}
\label{dn2}
\ee
Again, we have only displayed here the terms which will eventually contribute in our calculations. The complete list of operators is given in the Appendix B.

\begin{figure}
\centerline{
\resizebox{16cm}{1.2cm}{\includegraphics{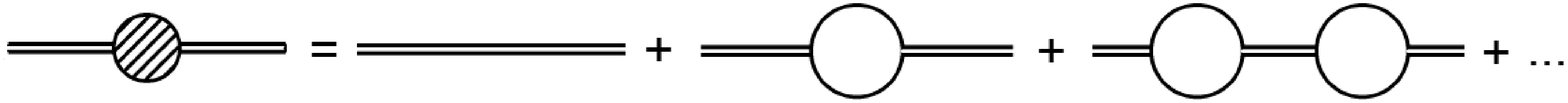}}}
\caption{{\footnotesize The dibaryon propagator gets an important contribution from resuming the bubble self-energy diagrams}}
\label{dbse}
\end{figure}

\begin{figure}
\centerline{
\resizebox{13.5cm}{1.3cm}{\includegraphics{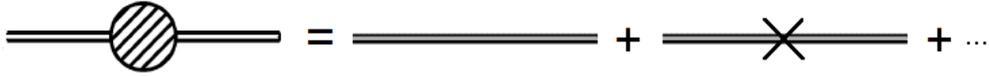}}}
\caption{\footnotesize{The LO dibaryon propagator for $p\gg \delta_{m_i}$ is the first term in the expansion of the full dibaryon propagator around $(-E+\d_{m_i})=0$. The second term is an effective vertex.}}
\label{propexp}
\end{figure}

As discussed in \cite{Soto:2007pg}, the dibaryon field propagator gets an important contribution to the self-energy due to the interaction with the nucleons (Fig.\ref{dbse}), which is always parametrically larger than the energy $E$. As a consequence the LO expression for the dibaryon field propagator becomes (in dimensional regularization (DR) and minimal subtraction (MS) scheme),
\be
\frac{i}{\d_{m_j}'+i\frac{A_j^2m_Np}{\pi}} \qquad j=s,v,
\label{dbself}
\ee
($p=\sqrt{m_N E}$) rather than the tree level expression $i/(-E+\delta_{m_j}'-i\eta)$. The size of the residual mass can be extracted computing the LO amplitude using the propagator (\ref{dbself}) and matching the result to the effective range expansion (ERE),

\be
\d_{m_i}' \sim \frac{1}{\p a_i}\lesssim \frac{m_{\p}^2}{\L_{\chi}} \qquad i=s,v.
\ee
$a_i$ are the scattering lengths. 
Therefore for $p \gg \frac{m_{\p}^2}{\L_{\chi}}$ the full propagator can be expanded. The first term of this expansion will be the LO propagator (Fig.\ref{propexp}),

\be
\frac{\pi }{A^2_i m_Np} \qquad i=s,v,
\label{dbexp}
\ee
the second term will be an effective vertex taking into account the effects due to $i(-E+\d_{m_i})$. Higher order terms in this expansion will be equivalent to multiple insertions of this vertex.
 
Furthermore for $p \gg \d_{m_i}'$ the LO Lagrangian becomes both scale and $SU(4)$ (spin-flavor Wigner symmetric) invariant, if the interactions with pions are neglected \cite{Mehen:1999qs}. Indeed, concerning $SU(4)$, the single nucleon sector is obviously invariant. Moreover, since all terms in (\ref{dbLO}) become subleading, one can redefine the dibaryon fields in such a way that all couplings in (\ref{dn}) are equal. In that case the dibaryon-nucleon interactions become $SU(4)$ invariant if the two dibaryon fields are chosen to form a $6^\ast$ representation of $SU(4)$. Scale invariance also holds because the dibaryon fields only appear in (\ref{dn}) and their scaling transformations can be chosen such that those terms are invariant.

Moreover equation (\ref{dbself}) implies that the dibaryon field should not be integrated out unless $p\ll \d_{m_i}'$, instead of $E\ll \d_{m_i}'$ as the tree level expression suggests. If $\d_{m_i}' \ll m_\pi$, it should also be kept as an explicit degree of freedom in the $\slashed{\p}$NNEFT, like in Refs. \cite{Ando:2004mm,Ando:2005dk,Ando:2005cz}. Nevertheless, if the dibaryon fields are integrated out, one can still organize the calculation in terms of nucleon fields by taking into acount suitable correlated enhancements in the local four nucleon interactions \cite{vanKolck:1998bw}. This is due to the fact that the path integral over dibaryon fields is Gaussian and can be carried out exactly. 

Except for the above mentioned contributions to the self-energy of the dibaryon fields, which become LO, the calculation can be organized perturbatively in powers of $1/\Lambda_\chi$. Hence one expects that any UV divergence arising in higher order calculations will be absorbed in a low energy constant of a higher dimensional operator built out of nucleon, dibaryon and pion fields (note that the linear divergence in the self-energy of the dibaryon fields due to the diagram in Fig.1b can be absorbed in $\delta_{m_i}$).

We shall restrict ourselves in the following to energies $E\lesssim m_\pi^2/\Lambda_\chi \ll m_\pi$, which implies nucleon three momenta $\sim m_\pi$. We shall follow the strategy of \cite{Eiras:2001hu}, which was inspired in the formalism of \cite{Pineda:1997bj}, and shall build a lower energy EFT, pNNEFT, with no explicit pion fields: the effects due to the pions will be encoded in the potentials (and redefinitions of the LECs). 
  
\section{The potential nucleon-nucleon effective theory \\with  dibaryon fields}

For energies $E\sim m_\pi^2/\Lambda_\chi \ll m_\pi$, the pion fields can be integrated out. This integration produces nucleon-nucleon potentials and redefinitions of low energy constants. Since $\delta_{m_i}'\sim m_\pi^2/\Lambda_\chi $ the dibaryon fields must be kept as explicit degrees of freedom in pNNEFT.

The Lagrangian in the $N_B=1$ sector reads
\be
\mathcal{L}_{N}= N^{\dag}\Bigl(i\partial_0-\d m_N+\frac{\vec{\partial}^2}{2m_N}+\frac{\vec{\partial}^4}{8m_N^3}\Bigr)N
\label{NpNN}
\ee

In the $N_B=2$ sector further two nucleon 
interactions (potentials) are induced. They read,
\be
\begin{split}
\mathcal{L}_{N N}=& 
\frac{1}{2}\int d^3{\bf r} N^{\dag}\s^\a\t^\r N(x_1)V_{\a\b ;\r \s}(x_1-x_2)N^{\dag}\s^\b\t^\s N(x_2) 
\end{split}
\label{pot}
\ee
$x_1^0=x_2^0=x^0$, ${\bf r}={\bf x}_1-{\bf x}_2$ and $x=(x_1+x_2)/2$
where $V_{\a\b ;\r \s}(x_1-x_2)$ is a generic potential ($\a ,\b ,\r ,\s =0,1,2,3$; $\;\;\t^0=\s^0=1$), which may be calculated in an expansion in $1/\L_\chi$ (in fact, beyond one loop it becomes an expansion in $\sqrt{1/\L_\chi}$ \cite{Mondejar:2006yu})

The terms with dibaryon fields and no nucleons read,
\be
\mathcal{L}'_{D}=D_{s,a}^{\dag}\Bigl(-i\pa_0+\delta_{m_s}\Bigr)D^a_s+
\vec{D}_v^{\dag}\Bigl(-i\pa_0+\delta_{m_v}\Bigr)\vec{D}_v
\label{dfp}
\ee
$\delta_{m_i}$, $i=s,v$ are the (redefined) dibaryon residual masses. 
Note that $\delta m_N$ in (\ref{NpNN}) 
can be reshuffled into $\delta_{m_i}$ by local field redefinitions. 
Note also that because of $\d_{m_i}'\ll m_\pi$ the quark mass dependence of $\d_{m_i}$ is a leading order effect.

The dibaryon-nucleon interactions remain the same as in (\ref{dn0}), except for the values of the $A_i$ which get modified.
\be
\begin{split}
\mathcal{L}_{DN}^{(1)}=&\frac{A_s}{\sqrt{2}}(N^{\dag}\s^2\t^a\t^2N^*)D_{s,a}+
\frac{A_s}{\sqrt{2}}(N^{\top}\s^2\t^2\t^aN)D^{\dag}_{s,a}+\\
&+\frac{A_v}{\sqrt{2}}(N^{\dag}\t^2\vec{\s}\s^2N^*)\cdot\vec{D}_v+\frac{A_v}{\sqrt{2}}(N^{\top}\t^2\s^2\vec{\s} N)\cdot\vec{D}_v^{\dag}
 \end{split}
\label{dnp}
\ee
\be
\begin{split}
\mathcal{L}_{DN}^{(2)}=&\frac{B_s}{\sqrt{2}}(N^{\dag}\s^2\t^a\t^2\pa^2N^*)D_{s,a}+\frac{B_s}{\sqrt{2}}(N^{\top}\s^2\t^2\t^a\pa^2N)D^{\dag}_{s,a}+
\\
&+\frac{B_v}{\sqrt{2}}(N^{\dag}\t^2\vec{\s}\s^2\pa^2N^*)\cdot\vec{D}_v+\frac{B_v}{\sqrt{2}}(N^{\top}\t^2\s^2\vec{\s}\pa^2 N)\cdot\vec{D}_v^{\dag}
\\
&+\frac{B'_v}{\sqrt{2}} (\pa_i N^{\dag}\t^2\s^i\s^2\pa_j N^*)D^j_v + \frac{B'_v}{\sqrt{2}}(\pa_i N^{\top}\t^2 \s^2 \s^i\pa_j N) D_v^{j\dag}
\end{split}
\label{dn2p}
\ee

The calculations in pNNEFT can be organized in ratios $E/p$ and $p/\Lambda_\chi$ (recall $m_\pi\sim p$). The UV divergences arising at higher orders will be absorbed by local terms build out of nucleon and dibaryon fields.

\section{Matching NNEFT to pNNEFT}

In the $N_B=1$
sector one loop pion contributions produce energy independent terms which are $O(m_\pi^2/\Lambda_\chi^2)$ \cite{Bernard:1993nj} and hence relevant for the N$^2$LO calculation, which together with 
the contribution $O(m_\pi/\Lambda_\chi)$ from terms proportional to the quark masses make up the nucleon residual mass $\d m_N$ in (\ref{NpNN}). 

In the $N_B=2$ sector 
the dibaryon residual masses
also get contributions $O(m_\pi/\Lambda_\chi)$ from terms proportional to the quark masses in (\ref{ordrep2}), and $O(m_\pi^2/\Lambda_\chi^2)$ ones from higher loop diagrams involving radiation pions, like the ones in fig.\ref{radpot}, fig.\ref{radpion}b and fig.\ref{yet},
\be
\begin{split}
&\d_{m_s}=\delta_{m_s}'+4m_q(s_1+s_2)+4A^2_s\frac{5}{3}\Bigl(\frac{g_A^2}{2f_{\p}^2}\Bigr)^2 \Bigl(\frac{m_N m_{\p}}{4\p}\Bigr)^3+\Bigl(\frac{g_A^2}{2f_{\p}^2}\Bigr)\frac{m^3_{\p}}{8\p}\frac{A_s^2}{A_v^2}+c_{sv}\Bigl(\frac{g_A}{f_{\p}^2}\Bigr)\frac{m^3_{\p}}{8\p}\frac{A_s}{A_v}\\
&\d_{m_v}=\delta_{m_v}'+4m_q v_1+4A^2_v\frac{5}{3}\Bigl(\frac{g_A^2}{2f_{\p}^2}\Bigr)^2 \Bigl(\frac{m_N m_{\p}}{4\p}\Bigr)^3+\Bigl(\frac{g_A^2}{2f_{\p}^2}\Bigr)\frac{m^3_{\p}}{8\p}\frac{A_v^2}{A_s^2}+c_{sv}\Bigl(\frac{g_A}{f_{\p}^2}\Bigr)\frac{m^3_{\p}}{8\p}\frac{A_v}{A_s}
\end{split}
\ee

\bfi
\centerline{
\begin{tabular}{cc}
\resizebox{12cm}{2cm}{\includegraphics{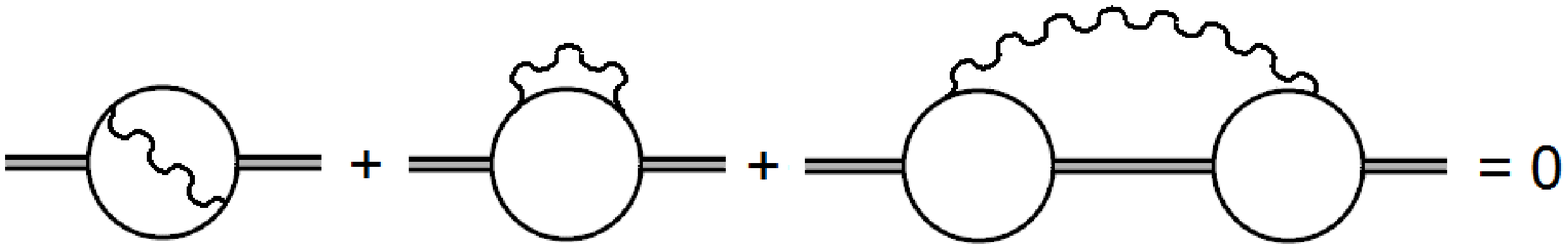}}& 
\resizebox{4.5cm}{2.1cm}{\includegraphics{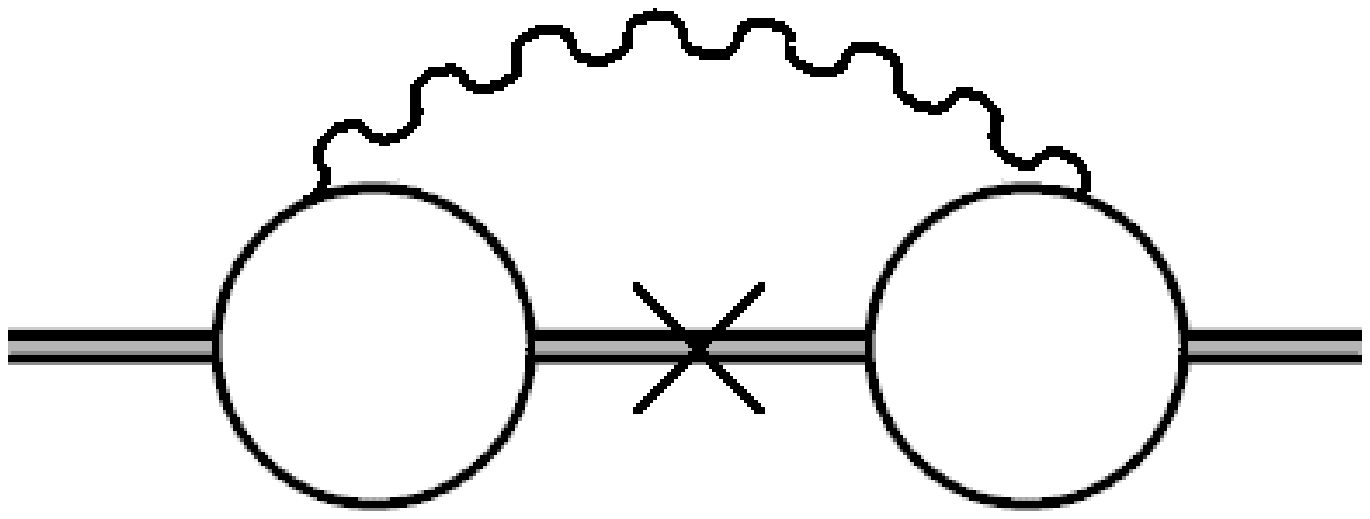}}
 \\
(a) & (b)
\end{tabular}
}
\caption{\footnotesize{ Order $O(m_\pi^{3/2}/\Lambda_\chi^{3/2})$ contributions to the dibaryon residual mass with one radiation pion. (a) These three diagrams sum up zero by Wigner symmetry. (b) Wigner symmetry is violated by (\ref{dbLO}), this is by insertions of $i(-E+\d_{m_i})$. Naively we would expect these diagrams to be of higher order, $O(m_\pi^{5/2}/\Lambda_\chi^{5/2})$, but the energy term is enhanced by the radiation pion up to $O(m_\pi^{2}/\Lambda_\chi^{2})$. Hence the cross in this diagram stands only for an insertion of the energy.}}
\label{radpion}
\efi

\bfi
\begin{minipage}[b]{0.65\linewidth}
\centerline{
\includegraphics[width=10cm]{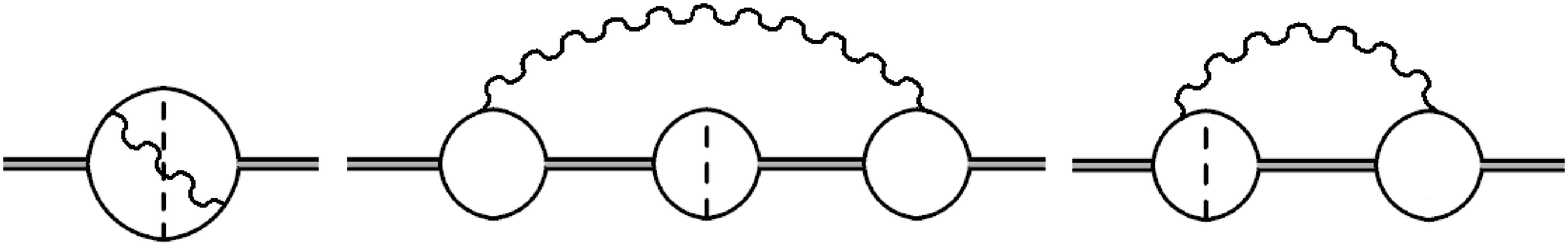}}
\caption{\footnotesize{Order $O(m_\pi^{2}/\Lambda_\chi^{2})$ contributions to the dibaryon residual mass with one radiation pion and one potential pion. Only diagrams with the potential pion inside the radiation pion loop contribute \cite{Fleming:1999ee}.}}
\label{radpot}
\end{minipage}
\hspace{0.1cm}
\begin{minipage}[b]{0.31\linewidth}
\centerline{\includegraphics[width=4cm]{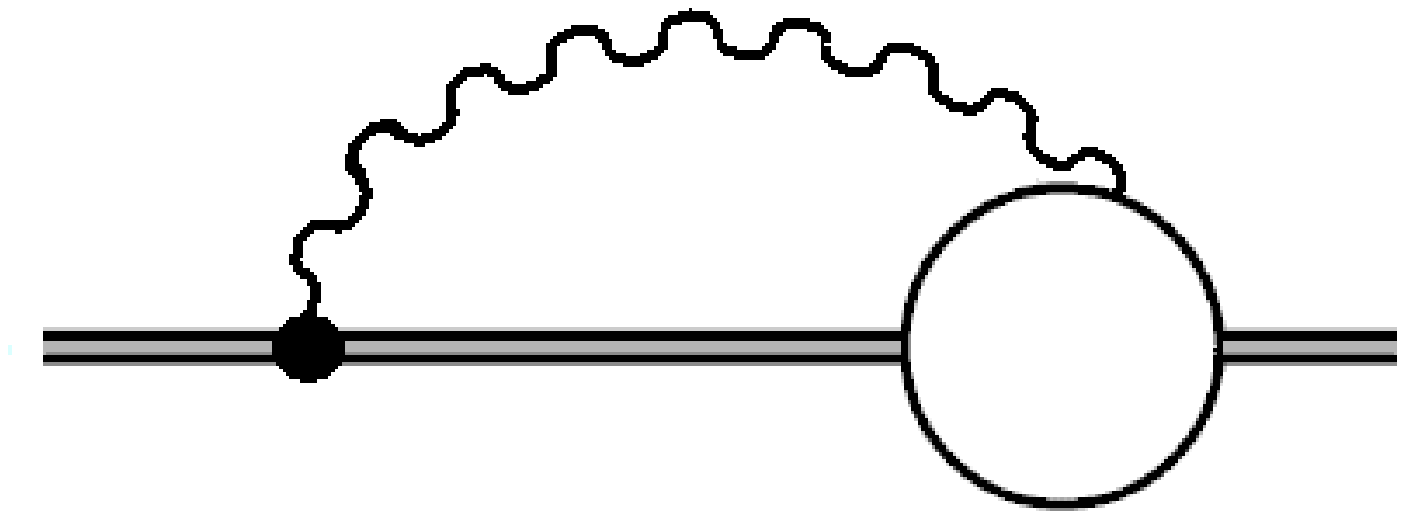}}
\caption{\footnotesize{Order $O(m_\pi^{2}/\Lambda_\chi^{2})$ contribution to the dibaryon residual mass involving the $c_{sv}$ vertex}}
\label{yet}
\end{minipage}
\efi

One may have also expected a contribution $O(m_\pi^{3/2}/\Lambda_\chi^{3/2})$ from diagrams in fig.\ref{radpion}a, but they add up to zero. This is not accidental, but a non-trivial consequence of Wigner symmetry \cite{Mehen:1999qs}. 

The dibaryon-nucleon vertices
may in principle get $O(m_\pi^2/\Lambda_\chi^2)$ from a pion loop, but they turn out to vanish, \
except for those which reduce to iterations of the OPE potential 
which will already be included in the calculations in pNNEFT and must not be considered in the matching. 
Note that for this to be so the matching calculation must be done according to the prescriptions of ref. \cite{Pineda:1998kn}, which we briefly recall in the Appendix A. This prescription gives results which differ from the on-shell prescription of ref. \cite{Fleming:1999ee} and are usually simpler. Agreement is eventually recovered at the level of physical amplitudes, in which a number of cancellations occur for the on-shell prescription. 
There is, however, a two loop contribution of this order involving radiation pions from the diagram in fig.\ref{ver2}.

\be
A_i\rightarrow A_i \biggl(1-4\frac{g^2_A m^2_{\pi}}{(4\pi f_{\pi})^2}\biggr)
\ee

\bfi
\centerline{
\resizebox{6.45cm}{2.22cm}{\includegraphics{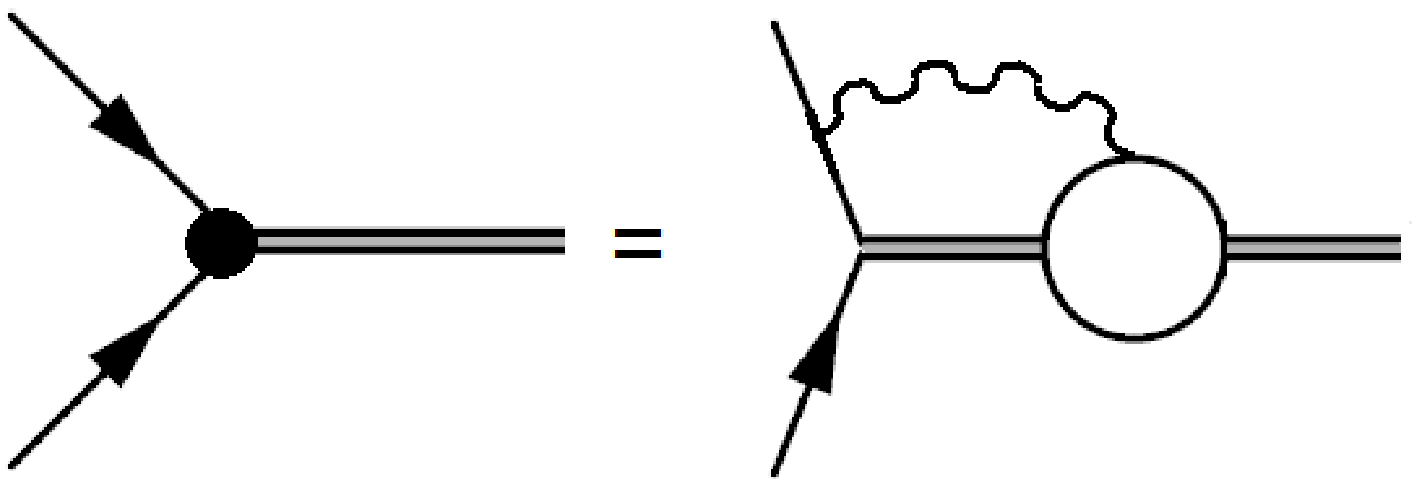}}}
\caption{\footnotesize{Matching of the effective vertex of the pNNEFT theory with the NNEFT vertex diagram.}}
\label{ver2}
\efi

Finally, in the two nucleon interactions (\ref{pot}), the one pion exchange is the only relevant contribution at this order, which produces the well known one pion exchange (OPE) potential,

\be
V_{\a\b,\r\s}(x_1-x_2)=-\frac{g^2_A}{2f^2_{\pi}}\int\frac{d^3q}{(2\pi)^3}\frac{q_\a q_\b}{\vec{q}^2+m^2_{\pi}}\d^{\r\s}e^{-i\vec{q}\cdot(\vec{x}_1-\vec{x}_2)}
\label{ope}
\ee
for $\a ,\b ,\r , \s =1,2,3$ and zero otherwise.

\section{Calculation in pNNEFT}

When $p\sim m_\pi$, we have already integrated out all higher energy and momentum scales in pNNEFT, and hence we already have the optimal EFT to carry out calculations. Moreover, for this momentum both the time derivative and the residual mass in the dibaryon Lagrangian are small and can be treated as $O(m_\pi/\Lambda_\chi)$ perturbations. 

Let us then focus on the calculation of nucleon-nucleon amplitudes up to N$^2$LO. At LO we get from fig.\ref{lo} the following S wave scale covariant Wigner symmetric amplitudes

\bfi
\centerline{\resizebox{4.5cm}{2.475cm}{\includegraphics{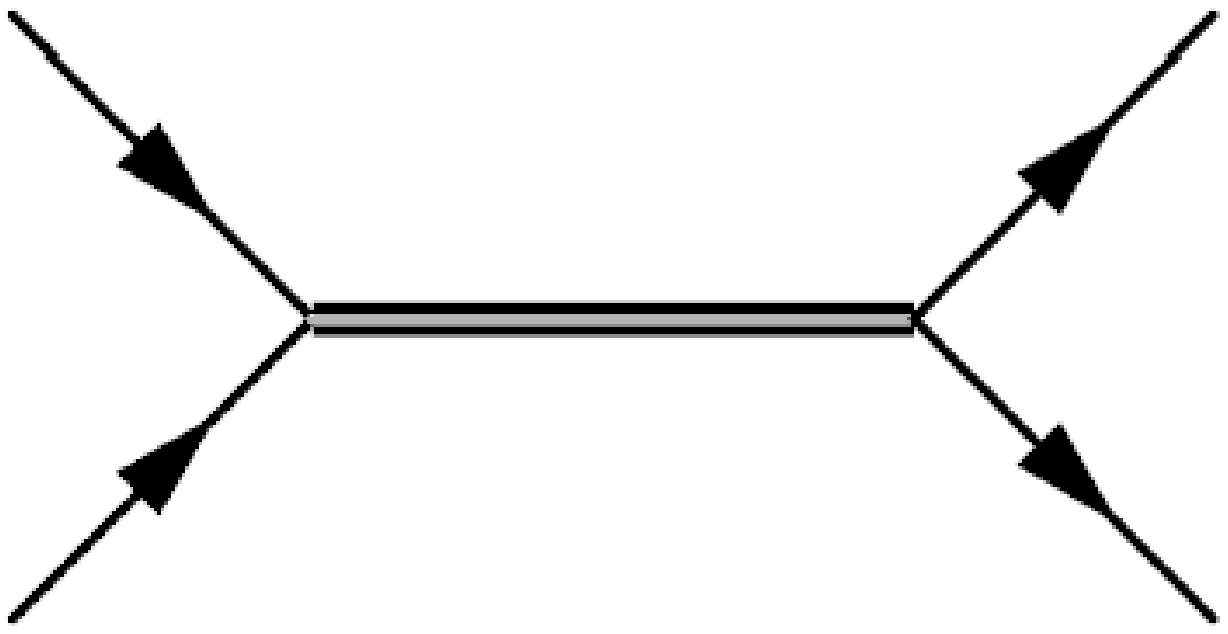}}}
\caption{\footnotesize{LO diagram}}
\label{lo}
\efi

\be
\mathcal{A}^{j}_{LO}=i\frac{4\pi }{m_Np} \qquad j=s,v 
\label{loa}
\ee  

At NLO we get from the diagrams in fig.\ref{nlo}a,

\bfi
\centerline{
\begin{tabular}{cc}
\resizebox{10.5cm}{2.22cm}{\includegraphics{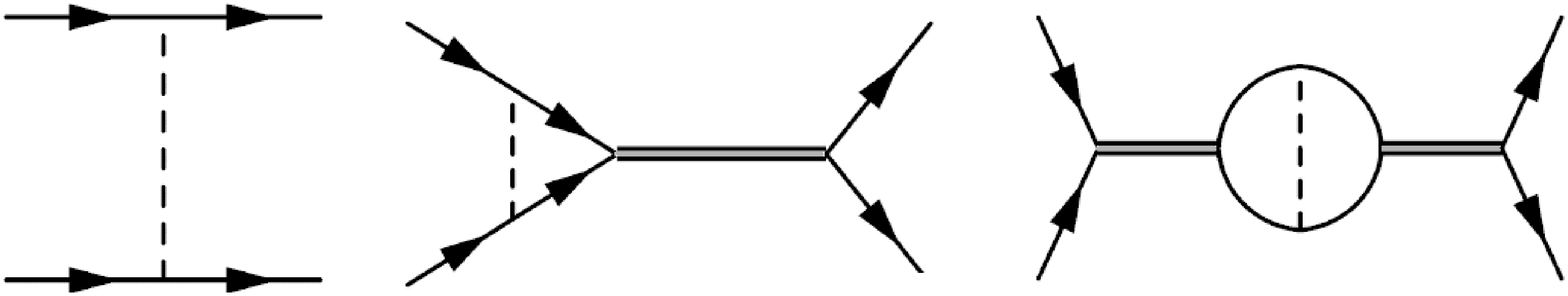}}& 
\resizebox{4cm}{2.22cm}{\includegraphics{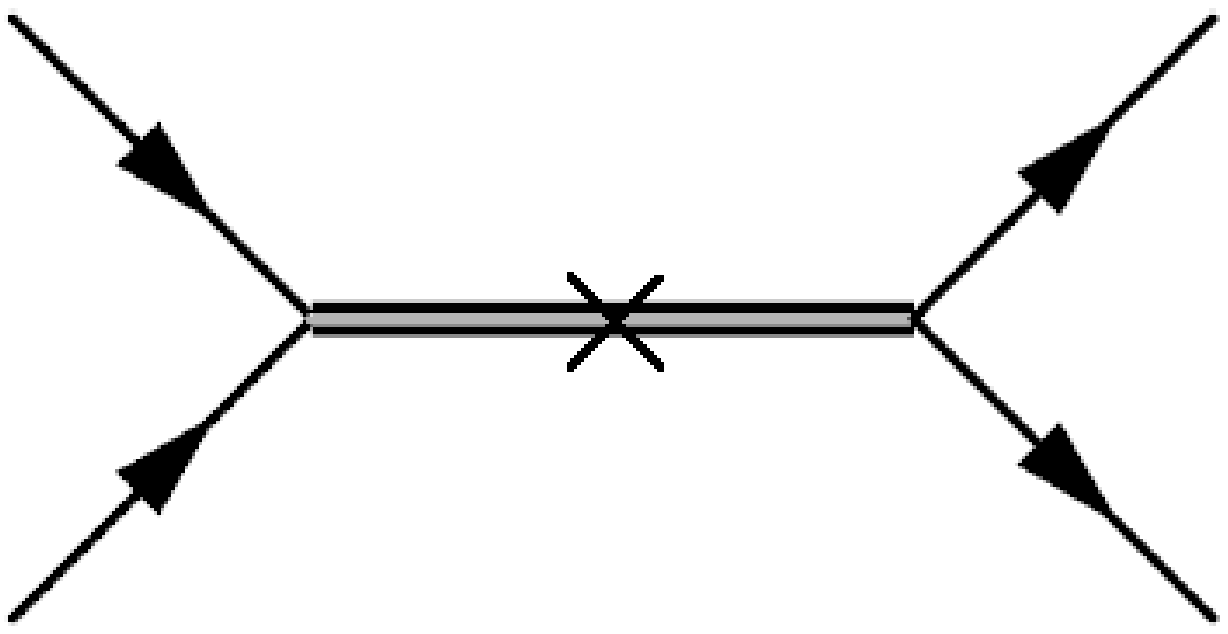}} \\
(a) & (b)
\end{tabular}
}
\caption{\footnotesize{(a) NLO diagrams with one potential pion exchange. (b) NLO diagram with one $i(-E+\d_{m_i})$ insertion}}
\label{nlo}
\efi

\be
\mathcal{A}^{i,I}_{NLO}=-\frac{g_A^2}{8f_{\pi}^2}\Bigl(\frac{m_N m_\pi}{4\pi}\Bigr)^2\ln\Bigl(1+\frac{4p^2}{m_{\pi}^2}\Bigr)(\mathcal{A}^{i}_{LO})^2  \qquad i=s,v 
\ee
and from the diagram in fig.\ref{nlo}b
\be
\mathcal{A}^{i,II}_{NLO}=\biggl(\frac{-E+\d_{m_i}}{4A^2_i}\biggr)(\mathcal{A}^{i}_{LO})^2 \qquad i=s,v
\ee
At N$^2$LO we obtain the following contributions. From the diagrams in fig.\ref{nnlopp},

\bfi
\centerline{
\resizebox{14cm}{4.5cm}{\includegraphics{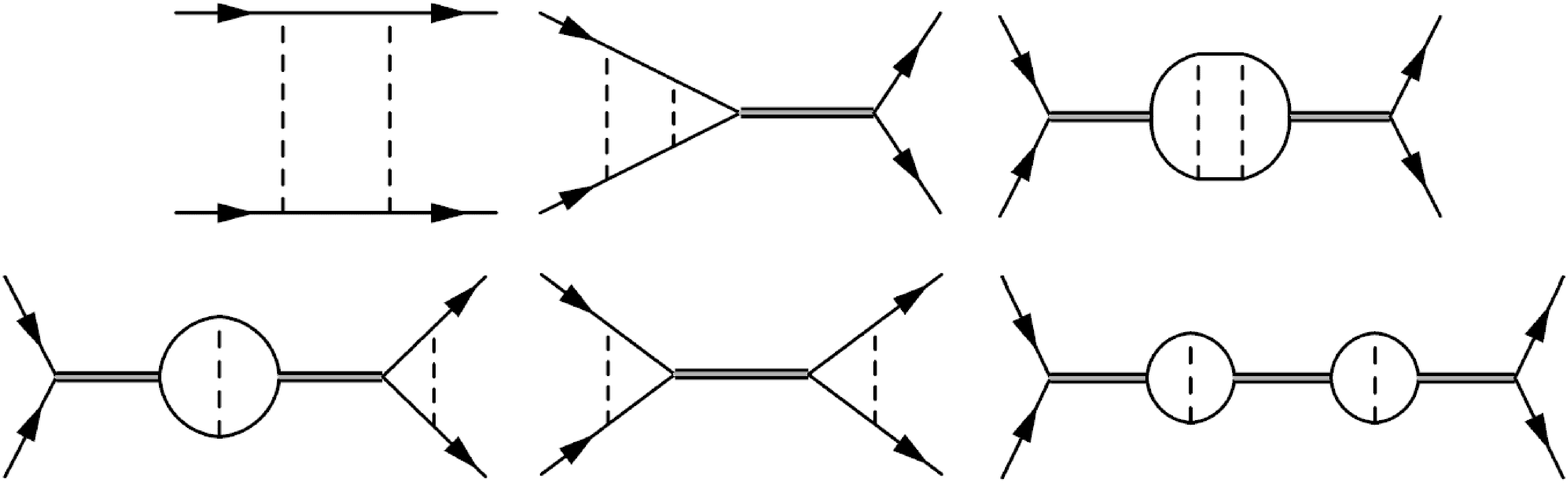}}}
\caption{\footnotesize{N$^2$LO diagrams with two potential pion exchange}}
\label{nnlopp}
\efi

\bfi
\centerline{
\resizebox{9cm}{2.22cm}{\includegraphics{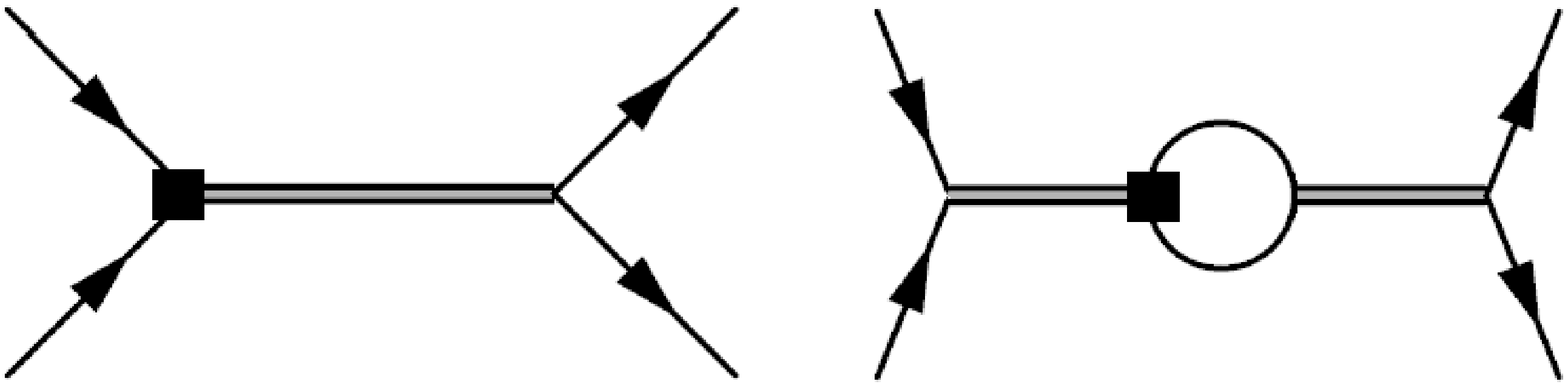}}}
\caption{\footnotesize{Diagrams with one vertex from (\ref{dn2}). 
}}
\label{ver1}
\efi

\bfi
\centerline{
\begin{tabular}{cc}
\resizebox{9cm}{2.22cm}{\includegraphics{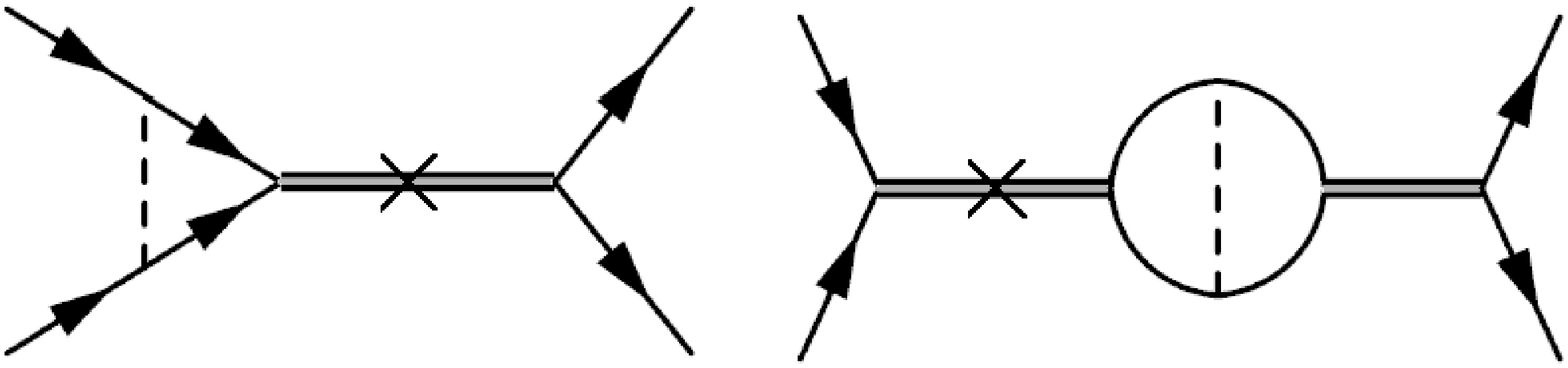}} &
\resizebox{4.5cm}{2.22cm}{\includegraphics{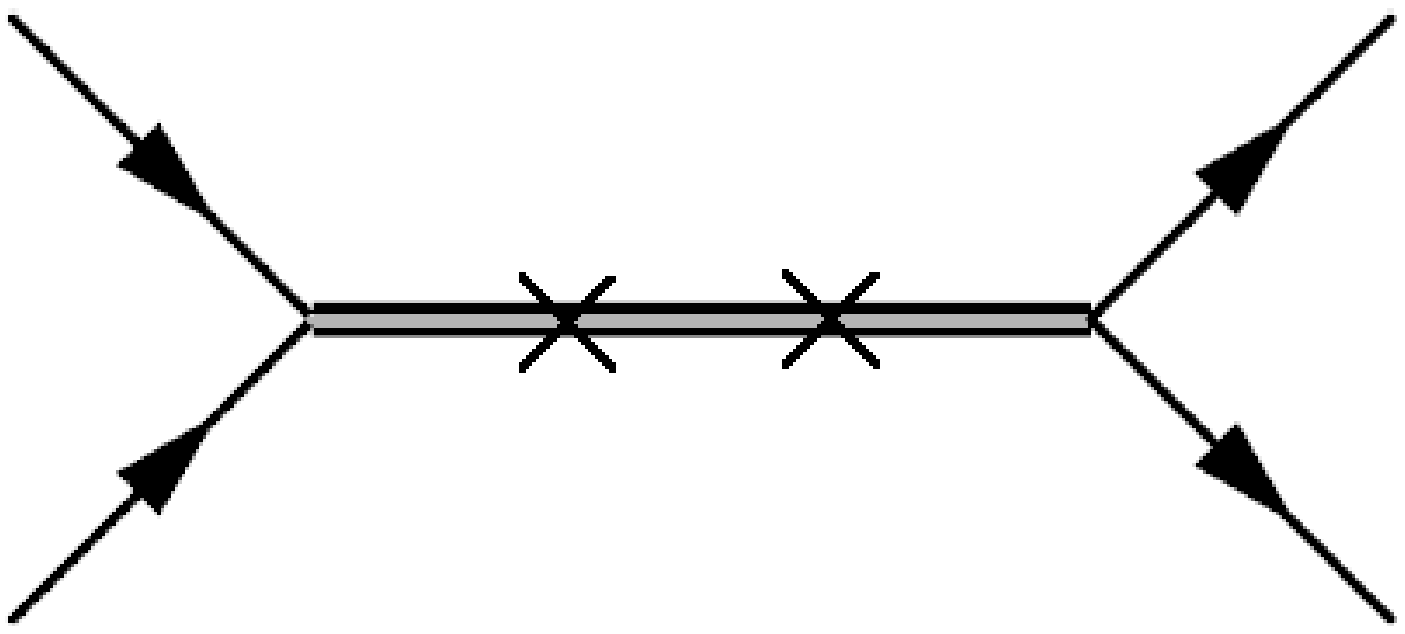}} \\
(a) & (b)
\end{tabular}
}
\caption{\footnotesize{N$^2$LO diagrams with, (a) one $i(-E+\d_{m_i})$ insertion and one potential pion, and (b) two $i(-E+\d_{m_i})$ insertions.}}
\label{nnlo2}
\efi

\be
\begin{split}
\mathcal{A}^{s,I}_{N^2LO}=&\Bigl(\frac{g_A^2}{2f_{\p}^2}\Bigr)^2 \Bigl(\frac{m_N m_{\p}}{4\p}\Bigr)^3 \Bigl(\frac{m_{\p}}{p}\Bigr) \Biggl[\frac{i}{16}\ln^2\Bigl(1+\frac{4p^2}{m^2_{\pi}}\Bigr)-\frac{1}{2}{\rm Im}\Bigl(Li_2\Bigl(\frac{-m_{\p}}{m_{\p}-2ip}\Bigr)\Bigr)+\\
&-\frac{1}{4}{\rm Im}\Bigl( Li_2\Bigl(\frac{m_{\p}+2ip}{-m_{\p}+2ip}\Bigr)\Bigr)+\frac{1}{2}\ln\Bigl(1+\frac{4p^2}{m^2_{\pi}}\Bigr)\arctan\Bigl(\frac{2p}{m_{\p}}\Bigr) \Biggr] (\mathcal{A}^{s}_{LO})^2
\end{split}
\ee

\be
\begin{split}
\mathcal{A}^{v,I}_{N^2LO}=&\Bigl(\frac{g_A^2}{2f_{\p}^2}\Bigr)^2 \Bigl(\frac{m_N m_{\p}}{4\p}\Bigr)^3\Biggl[6\Bigl(\frac{p}{m_{\p}}\Bigr)^2-\frac{3}{4}\Bigl(\frac{m_{\p}}{p}\Bigr)^2+4-\Bigl(\frac{9}{4}\Bigl(\frac{m_{\p}}{p}\Bigr)^4+3\Bigl(\frac{m_{\p}}{p}\Bigr)^2\Bigr)\ln 2+\\
&+\frac{i}{16}\Bigl(\frac{m_{\p}}{p}\Bigr)\ln^2\Bigl(1+\frac{4p^2}{m^2_{\pi}}\Bigr)+i\biggl(\frac{9}{8}\Bigl(\frac{m_{\p}}{p}\Bigr)^3-\Bigl(\frac{3}{2}\Bigl(\frac{m_{\pi}}{p}\Bigr)^2+\frac{9}{8}\Bigl(\frac{m_{\pi}}{p}\Bigr)^4\Bigr)\arctan\Bigl(\frac{2p}{m_{\p}}\Bigr)+\\
&+\Bigl(\frac{m_{\p}}{2p}+\frac{9}{32}\Bigl(\frac{m_{\p}}{p}\Bigr)^5+\frac{3}{4}\Bigl(\frac{m_{\p}}{p}\Bigr)^3\Bigr)\arctan^2\Bigl(\frac{2p}{m_{\p}}\Bigr)\biggr)+\frac{m_{\p}}{4p}\ln\Bigl(1+\frac{4p^2}{m^2_{\p}}\Bigr)\arctan\Bigl(\frac{2p}{m_{\p}}\Bigr)+\\
&+\frac{3}{4}\Bigl(\frac{3}{4}\Bigl(\frac{m_{\p}}{p}\Bigr)^4+\Bigl(\frac{m_{\p}}{p}\Bigr)^2 \Bigr)\ln\Bigl(1+\frac{4p^2}{m^2_{\p}}\Bigr)-\frac{3}{4}\Bigl(\Bigl(\frac{m_{\p}}{p}\Bigr)^2+\frac{3}{4}\Bigl(\frac{m_{\p}}{p}\Bigr)^4\Bigr)\ln\Bigl(1+\frac{p^2}{m_{\p}^2}\Bigr)+\\
&+\Bigl(6\Bigl(\frac{p}{m_{\p}}\Bigr)^3+\frac{6p}{m_{\p}}-\frac{3m_{\p}}{4p}-\frac{9}{8}\Bigl(\frac{m_{\p}}{p}\Bigr)^3\Bigr)\arctan\Bigl(\frac{p}{m}\Bigr)-\frac{3}{4}\Bigl(\frac{3}{8}\Bigl(\frac{m_{\p}}{p}\Bigr)^5+\Bigl(\frac{m_{\p}}{p}\Bigr)^3+\frac{m_{\p}}{p}\Bigr) \\
&\biggl(2 {\rm Im}\Bigl(Li_2\Bigl(\frac{-m_{\p}}{m_{\p}-2ip}\Bigr)\Bigr)+{\rm Im}\Bigl(Li_2\Bigl(\frac{m_{\p}+2ip}{-m_{\p}+2ip}\Bigr)\Bigr)-\ln\Bigl( 1+\frac{4p^2}{m^2_{\p}}\Bigr)\arctan\Bigl(\frac{2p}{m_{\p}}\Bigr)\biggr)\Biggr] (\mathcal{A}^{v}_{LO})^2
\end{split}
\ee

The sum of the diagrams in fig.\ref{ver1} 
turns out to be zero for the $^1S_0$ and $^3S_1$ channels (they only contribute to the $^3S_1$-$^3D_1$ mixing, see below). This can be understood as follows: these diagrams involve corrections to the nucleon-dibaryon vertices of order $(\frac{m_{\p}}{\L_{\chi}})^2 $. We can redefine the dibaryon fields in order to remove these corrections from nucleon-dibaryon vertices, as a consequence these corrections would appear in the $N_B=2$ sector (\ref{NpNN}), however, since this sector is subleading, the new operators induced by the field redefinition in this sector are of higher order.

From the diagrams in fig.\ref{nnlo2}
\be
\begin{split}
\mathcal{A}^{j,II}_{N^2LO}=&-\biggl(\frac{-E+\d_{m_j}}{4A^2_j}\biggr)\Bigl(\frac{g^2_A}{2f_{\p}^2}\Bigr) \Bigl(\frac{m_Nm_{\p}}{4\p}\Bigr)\frac{m_{\p}}{p}\biggl(\arctan\Bigl(\frac{2p}{m_{\p}} \Bigr)+\frac{i}{2}\ln\Bigl(1+\frac{4p^2}{m^2_{\p}}\Bigr) \biggr)(\mathcal{A}^{j}_{LO})^2+\\
&+\biggl(\frac{-E+\d_{m_j}}{4A^2_j}\biggr)^2(\mathcal{A}^{j}_{LO})^3 \qquad j=s,v
\end{split}
\label{annlo2}
\ee

Finally there are two contributions coming from relativistic corrections. The first one comes from using $i/(p^0-\frac{p^2}{2m_N}+\frac{p^4}{8m^3_N}+i\e)$ instead of $i/(p^0-\frac{p^2}{2m_N}+i\e)$ as the nucleon propagator. We obtain the contribution in Fig.\ref{relcor},

\be
\mathcal{A}^{j,a}_{N^2LO}=i\biggl(\frac{5p^3}{32\pi m_N}\biggr)(\mathcal{A}^{j}_{LO})^2 \qquad j=s,v.
\ee

\bfi
\centerline{
\resizebox{6cm}{2.75cm}{\includegraphics{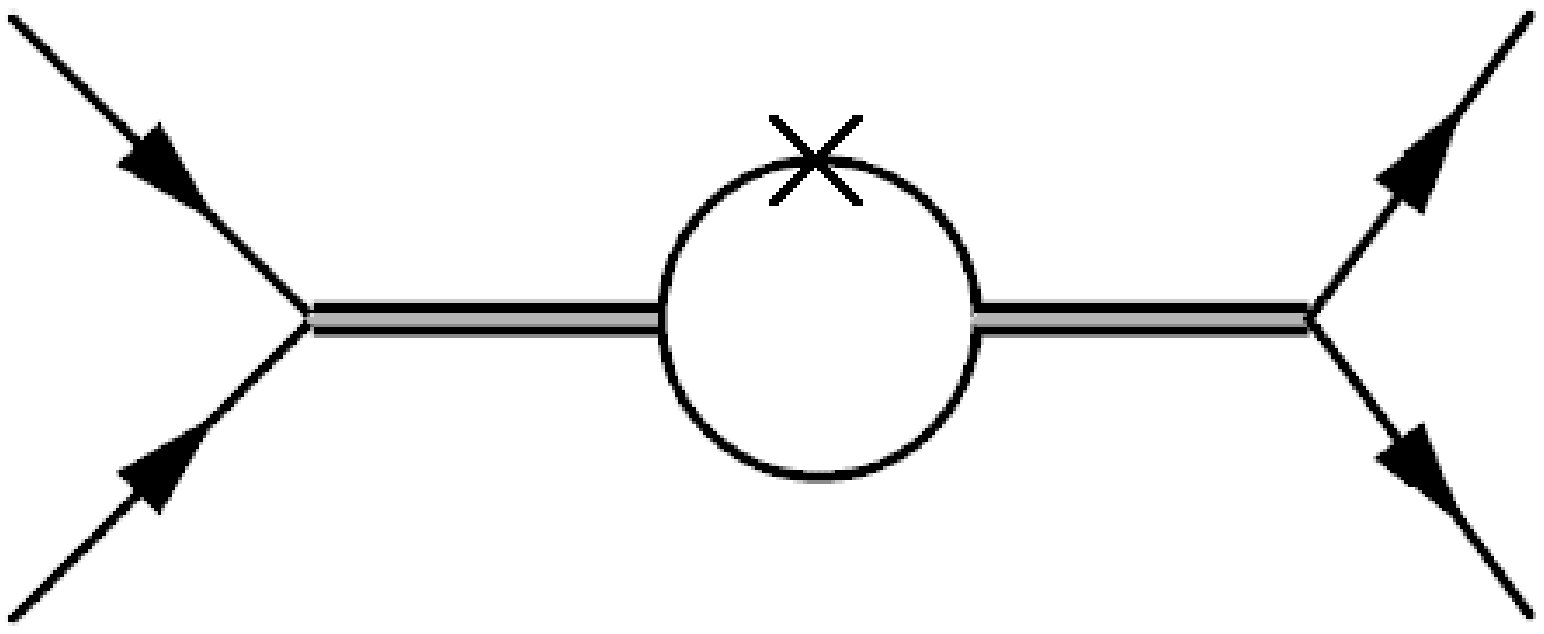}}}
\caption{\footnotesize{The cross in one of the nucleon propagators stands for the use of the relativistic correction. In order to compute this diagram the relativistic correction has been expanded up to first order}}
\label{relcor}
\efi

Another contribution arise when using the first relativistic correction to the dispersion relation of the nucleons,

\be 
p^0=\frac{p^2}{2m_N}-\frac{p^4}{8m^3_N},
\ee 
in the bubble self-energy diagram of Fig.\ref{dbse}. This results in the following contribution to the amplitude,
 
\be
\mathcal{A}^{j,b}_{N^2LO}=-i\biggl(\frac{p^3}{32\pi m_N}\biggr)(\mathcal{A}^{j}_{LO})^2 \qquad j=s,v.
\ee

However a new normalization of the amplitude that takes into account the new dispersion relation has to be considered,

\be
S=1+\frac{i}{2\p}\frac{p^2}{dE/dp}\mathcal{A}=1+i\frac{m_Np}{2\p}\biggl(1+\frac{p^2}{2m^2_N}\biggr)\mathcal{A},
\ee
this new normalization induces a new contribution to the $S$ matrix, $i\frac{p^3}{4\pi m_N}\mathcal{A}^{j}_{LO}$, that exactly cancels the contributions coming from the aforementioned relativistic contributions to the amplitude.

\paragraph{}The $^3S_1$-$^3D_1$ mixing amplitude has no contribution at LO. At NLO the first two diagrams of Fig.\ref{nlo}a are the only contribution,

\be
\mathcal{A}^{mix}_{NLO}=\sqrt{2}\Bigl(\frac{g^2_A}{2f^2_{\p}}\Bigr)\frac{m_Nm_{\p}}{4\p}\Bigl[-\frac{3}{4}\Bigl(\frac{m_{\p}}{p}\Bigr)^2+\Bigl(\frac{m_{\p}}{2p}+\frac{3}{8}\Bigl(\frac{m_{\p}}{p}\Bigr)^3 \Bigr)\arctan\Bigl(\frac{2p}{m_{\p}}\Bigr) \Bigr]\mathcal{A}^{v}_{LO}.
\label{mixnlo}
\ee

At N$^2$LO diagrams of Fig.\ref{nnlopp} with one (or two) potential pion 
exchange in the nucleon external legs
give the following contribution,

\be
\begin{split}
\mathcal{A}^{mix,I}_{N^2LO}=&\sqrt{2}\Bigl(\frac{g^2_A}{2f^2_{\p}}\Bigr)^2\Bigl(\frac{m_Nm_{\p}}{4\p}\Bigr)^2\Bigl[ \mathcal{Z}\Bigl(\frac{p}{m_{\p}}\Bigr)-i\frac{3p}{2m_{\p}}\mathcal{Y}\Bigl(\frac{p}{m_{\p}}\Bigr)+\\
&-i\Bigl(\frac{p}{m_{\p}}+\frac{m_{\p}}{2p}\ln\Bigl(1-i\frac{2p}{m_{\p}}\Bigr)\Bigr)\mathcal{X}\Bigl(\frac{p}{m_{\p}}\Bigr)\Bigr]\mathcal{A}^v_{LO}.
\end{split}
\ee

The ${\cal X}$, ${\cal Y}$, ${\cal Z}$ functions where defined in \cite{Fleming:1999bs} and we write them here for completeness,

\begin{eqnarray}
{\cal X}(\alpha) &=& -\frac{3}{4\alpha^2}-\frac{3i}{4\alpha}+\frac{i\alpha}2 + 
     i \bigg( \frac1{2\alpha} + \frac3{8\alpha^3} \bigg) \ln(1-2 i\alpha) \,, \\
{\cal Y}(\alpha) &=& -\frac25 + \frac3{10\alpha^2} + \bigg(\frac{3}{8\alpha^5} + 
      \frac{5}{4\alpha^3} - \frac{2\alpha}{5} \bigg) \tan^{-1}(\alpha)  
      -\bigg(\frac{3}{8\alpha^5} +\frac{5}{4\alpha^3} \bigg) \tan^{-1}(2\alpha) \\*
   && + \frac{(15-4\alpha^2)}{80\alpha^6} \ln(1+\alpha^2)  
     - \frac{(3+16\alpha^2+16\alpha^4)}{32\alpha^7} {\rm Im}\,\bigg[ Li_2\Big( 
      \frac{2\alpha^2+i\alpha}{1+4\alpha^2} \Big) + Li_2( -2\alpha^2-i\alpha )
      \bigg]  \nn\\*
   && + i \bigg[ \frac{3}{8\alpha^3} +\frac{1}{2\alpha} -\frac{\alpha}{2} - 
      \frac{(3+10\alpha^2)}{16\alpha^5}  \ln(1+4\alpha^2)
      +\frac{(3+16\alpha^2+16\alpha^4)}{128\alpha^7} \ln^2(1+4\alpha^2) \bigg]\,, \nn
     \nn\\ 
{\cal Z}(\alpha) &=& -\frac{7}{40} +\frac{9i}{16\alpha^3}+\frac{21}{40\alpha^2} +
  \frac{3i}{40\alpha}-\frac{3i\alpha}{5} + \frac{29\alpha^2}{200} + 
  \Big(\frac{3\alpha^2}{5} -\frac{9}{16\alpha^4}-\frac{15}{8\alpha^2} \Big) \ln{2} \\*
 && + \frac{3\,(16\alpha^7-50\alpha^3-4i\alpha^2-15\alpha+15i)}{80\alpha^5} 
    \ln(1-i\alpha) \nn\\*
 && + \frac{(-9\,i + 27\,\alpha  - 24\,i\,{{\alpha }^2} + 
    78\,{{\alpha }^3} - 16\,{{\alpha }^5})}{32\alpha^5} \ln(1-2i\alpha) \nn \\*
 && - \frac{(9+48\alpha^2+48\alpha^4)}{64\alpha^6} \bigg[ \frac32 \ln^2(1-2i\alpha)
    + 2 Li_2(-1+2i\alpha) + Li_2 \Big(\frac{1+2i\alpha}{-1+2i\alpha}\Big) +\frac{\pi^2}4
    \bigg]   \nn \,.
\end{eqnarray}

The derivative vertex of (\ref{dn2}) proportional to $B_v'$ contributes to the mixing amplitude through the first diagram of Fig.\ref{ver1},

\be
\mathcal{A}^{mix,II}_{N^2LO}=i\sqrt{2}p^2\frac{B'_v}{A_v}\mathcal{A}^v_{LO}.
\ee

The last contribution to the $^3S_1$-$^3D_1$ mixing amplitude comes from the first diagram of Fig\ref{nnlo2}a,
 
\be
\begin{split}
\mathcal{A}^{mix,III}_{N^2LO}=&\sqrt{2}\Bigl(\frac{g^2_A}{2f^2_{\p}}\Bigr)\frac{m_Nm_{\p}}{4\p}\Bigl(\frac{-E+\d_{m_v}}{4A_v^2}\Bigr)\Bigl[-\frac{3}{4}\Bigl(\frac{m_{\p}}{p}\Bigr)^2+\Bigl(\frac{3}{8}\Bigl(\frac{m_{\p}}{p}\Bigr)^3+\frac{m_{\p}}{2p}\Bigr)\arctan\Bigl(\frac{2p}{m_{\p}}\Bigr)+\\
&+i\Bigl\{-\frac{3m_{\p}}{4p}+\frac{p}{2m_{\p}}+\Bigl(\frac{m_{\p}}{4p}+\frac{3}{16}\Bigl(\frac{m_{\p}}{p}\Bigr)^3\Bigr)\ln\Bigl(1+\frac{4p^2}{m^2_{\p}}\Bigr) \Bigr\} \Bigr](\mathcal{A}^{v}_{LO})^2.
\end{split}
\ee

\paragraph{}The $^3D_1$ amplitude starts at NLO with the contribution coming from the one pion exchange diagram,

\be
\begin{split}
\mathcal{A}^{^3D_1}_{NLO}=\Bigl(\frac{g^2_A}{2f^2_{\p}}\Bigr)\Bigl[-\frac{1}{2}-\frac{3}{4}\Bigl(\frac{m_{\p}}{p}\Bigr)^2+\Bigl(\frac{3}{16}\Bigl(\frac{m_{\p}}{p}\Bigr)^4+\frac{1}{2}\Bigl(\frac{m_{\p}}{p}\Bigr)^2\Bigr)\ln\Bigl(1+\frac{4p^2}{m^2_{\p}}\Bigr)\Bigr].
\end{split}
\ee

At N$^2$LO there are two contributions from Fig.\ref{nnlopp}
from the two diagrams in which all external nucleon legs have a potential pion exchange.
The corresponding amplitudes are,

\be
\begin{split}
\mathcal{A}^{^3D_1,I}_{N^2LO}=&\Bigl(\frac{g^2_A}{2f^2_{\p}}\Bigr)^2\Bigl(\frac{m_Nm_{\p}}{4\p}\Bigr)\frac{3}{2}\biggl(\\
&-\frac{2}{7}+\frac{54}{35}\Bigl(\frac{m_{\pi}}{p}\Bigr)^4-\frac{19}{70}\Bigl(\frac{m_{\p}}{p}\Bigr)^2+\Bigl(\frac{9}{8}\Bigl(\frac{m_{\p}}{p}\Bigr)^5+\frac{7}{4}\Bigl(\frac{m_{\p}}{p}\Bigr)^3+\frac{4m_{\p}}{5p}-\frac{2 p}{7 m_{\p}}\Bigr)\arctan\Bigl(\frac{p}{m_{\p}}\Bigr)+\\
&-\Bigl(\frac{9}{8}\Bigl(\frac{m_{\p}}{p}\Bigr)^5+\frac{7}{4}\Bigl(\frac{m_{\p}}{p}\Bigr)^3\Bigr)\arctan\Bigl(\frac{2 p}{m_{\p}}\Bigr)-\Bigl(\frac{549}{560}\Bigl(\frac{m_{\p}}{p}\Bigr)^6+\frac{3}{4}\Bigl(\frac{m_{\p}}{p}\Bigr)^4\Bigr) \ln\Bigl(1+\frac{p^2}{m_{\p}^2}\Bigr)+\\
&-\Bigl(\frac{9}{32}\Bigl(\frac{m_{\p}}{p}\Bigr)^7+\Bigl(\frac{m_{\p}}{p}\Bigr)^5+\Bigl(\frac{m_{\p}}{p}\Bigr)^3\Bigr) {\rm Im}\Bigl[Li_2\Bigl(\frac{-i m_{\p} p-2 p^2}{m_{\p}^2}\Bigr)+Li_2\Bigl(\frac{i m_{\p} p+2 p^2}{m_{\p}^2+4p^2}\Bigr)\Bigr]+\\
&+i \biggl\{\frac{9}{8}\Bigl(\frac{m_{\p}}{p}\Bigr)^3-\frac{m}{2p}+\frac{p}{2m}-\Bigl(\frac{9}{16}\Bigl(\frac{m_{\p}}{p}\Bigr)^5+\frac{7}{8}\Bigl(\frac{m_{\p}}{p}\Bigr)^3\Bigr)\ln\Bigl(1+\frac{4p^2}{m_{\p}^2}\Bigr)\\
&+\Bigl(\frac{9}{128}\Bigl(\frac{m_{\p}}{p}\Bigr)^7+\frac{1}{4}\Bigl(\frac{m_{\p}}{p}\Bigr)^5+\frac{1}{4}\Bigl(\frac{m_{\p}}{p}\Bigr)^3\Bigr)\ln^2\Bigl(1+\frac{4p^2}{m^2}\Bigr)\biggr\}\biggr)
\end{split}
\ee

\be
\begin{split}
\mathcal{A}^{^3D_1,II}_{N^2LO}= 2 \Bigl(\frac{g^2_A}{2f^2_{\p}}\Bigr)^2\Bigl(&\frac{m_Nm_{\p}}{4\p}\Bigr)^2\Bigl[-\frac{3}{4}\Bigl(\frac{m_{\p}}{p}\Bigr)^2+\Bigl(\frac{m_{\p}}{2p}+\frac{3}{8}\Bigl(\frac{m_{\p}}{p}\Bigr)^3\Bigr)\arctan\Bigl(\frac{2p}{m_{\p}}\Bigr)+\\
&+i\Bigl\{-\frac{3m_{\p}}{4p}+\frac{p}{2m_{\p}}+\Bigl(\frac{m_{\p}}{4p}+\frac{3}{16}\Bigl(\frac{m_{\p}}{p}\Bigr)^3\Bigr)\ln\Bigl(1+\frac{4p^2}{m^2_{\p}}\Bigr)\Bigr\}\Bigr]^2\mathcal{A}^{v}_{LO}.
\end{split}
\ee

\section{The pionless nucleon-nucleon effective field theory}
\label{npeft}
For $p\lesssim \frac{m^2_\p}{\L_{\chi}}$ the calculation must be organized in a different way. This is very much facilitated if we integrate out nucleon three momenta of the order of $m_\pi$ first, which leads to the so called pionless nucleon-nucleon EFT ($\slashed{\p}$NNEFT) \cite{Kaplan:1996xu,vanKolck:1998bw,Beane:2000fi}. This EFT has been successfully used in numerous processes at very low energy (see \cite{Epelbaum:2008ga} for a recent review). The Lagrangian of the $N_B=1$ sector of this theory remains the same as in pNNEFT (\ref{NpNN}) (the relativistic correction becomes negligible). For the $N_B=2$ sector the only formal difference from pNNEFT is that the non-local potentials ({\ref{pot}) become local and can be organized in powers of $p^2/m_\pi^2$. The OPE potential in (\ref{ope}) becomes $O(p^2/m_\pi^2\Lambda_\chi^2)$ and hence beyond N$^3$LO in this
region. The derivative dibaryon-nucleon vertices in (\ref{dn2p}) also become beyond this order. The remaining terms in the Lagrangian are the same as those in pNNEFT, namely (\ref{dfp}) and (\ref{dnp}), with the parameters redefined as follows. Diagrams in Fig.\ref{nlo}a and Fig.\ref{nnlopp} containing one (or two) potential pion inside a nucleon bubble will contribute to the dibaryon time derivative term as well as the dibaryon residual mass. Contributions to the dibaryon time derivative can be reabsorbed by field redefinitions of dibaryon fields, while contributions to the residual mass simply redefine it. The dibaryon-nucleon vertex (\ref{dnp}) gets contributions from diagrams containing one (or two) potential pion in the dibaryon-nucleon vertex, redefining the $A_i$. There are also higher order self-energy diagrams for the dibaryon fields not shown in paper which contribute to the redefinitions of the residual dibaryon masses at the order we are interested in, like the ones involving three OPE or the two pion exchange potential in a nucleon bubble. We will chose to  reshuffle all matching contributions to the dibaryon-nucleon vertices to the residual masses through field redefinitions of the dibaryon fields. This way the coupling constants $A_i$ will remain the same as in pNNEFT while all the new dependences are carried by the residual masses.

\paragraph{}Since the dibaryon residual masses are no longer small, but of the same order, when compared to $p$, residual masses have to be kept in the dibaryon propagators. Hence we will use (\ref{dbself}) rather than (\ref{dbexp}) as the dibaryon propagator.
 
The LO amplitude for the pionless EFT is obtained from the diagram in Fig.\ref{lo} using the new dibaryon propagator,

\be
\mathcal{A}^{j,\slashed{\p}}_{LO}=\frac{-4A^2_j}{\d_{m_j}+i\frac{A^2_jm_Np}{\p}} \qquad j=s,v.
\ee

Note that the LO amplitude is of order $O(1/m_\pi^2)$ instead of $O(1/m_\pi\Lambda_\chi)$ as in pNNEFT, however since contributions to the S matrix are proportional to the momentum the final size of the LO contributions to the observables (as well as the NLO and N$^2$LO ones) remains the same 
as in the high energy region. Note also that both scale invariance and Wigner symmetry are lost in the low energy region.

The form of the amplitude remains the same at NLO (only $A_i$ and $\delta_{m_i}$ get redefined). At N$^2$LO (i.e. $O(1/\Lambda_\chi^2)$) a contribution corresponding to Fig.\ref{nlo}b
arises, 

\be
\mathcal{A}^{i,\slashed{\p}}_{N^2LO}=-\biggl(\frac{E}{4A^2_i}\biggr)(\mathcal{A}^{i,\slashed{\p}}_{LO})^2 \qquad i=s,v.
\ee

The form of the N$^2$LO expression turns out to be valid also up to N$^3$LO (i.e. $O(m_{\p}/\Lambda_\chi^3)$, again only $A_i$ and $\delta_{m_i}$ get redefined)
The sum of diagrams in Fig.\ref{ver1}. is no longer zero but the momentum dependence of the vertex involved makes them beyond N$^3$LO.

\paragraph{}No contributions to $\mathcal{A}^{mix,\slashed{\p}}$ or to $\mathcal{A}^{^3D_1,\slashed{\p}}$ appear up to N$^2$LO (the first diagram of Fig.\ref{nnlo2}.a contributes to $\mathcal{A}^{mix,\slashed{\p}}$ at N$^3$LO; this amplitude matches a straightforward expansion for $p\ll m_{\p}$ of the pNNEFT mixing amplitude).

\section{The $^1 S_0$ channel}

\paragraph{}In this section we compare our results for the $^1 S_0$ channel with its corresponding phase shift data. In order to compute the phase shift the amplitude has been introduced in $\exp(2i\d)=1+ip m_N\mathcal{A}/2\p$. After expanding both sides in powers of $(m_{\p}/\L_{\chi})^n$ the expressions for $\d^{LO}$, $\d^{NLO}$ and, $\d^{N^2LO}$ are obtained.

We will not display the results for $\d^{LO}$. At this order our approach does not uniquely determine the phase shift in the high energy region. This can be easily seen if the expression for the phase shift is expressed in terms of the real and imaginary parts of the amplitude,

\be
\d=\arctan\Bigl(\frac{{\rm Im}\mathcal{A}}{{\rm Re}\mathcal{A}}\Bigr).
\ee
Since our LO amplitude (\ref{loa}) has no real part, then $\delta=\pm \pi /2$. Continuity with the low energy expression selects the plus sign.

$A_s$ and $\d_{m_s}$ receive  corrections in the matching from NNEFT to pNNEFT, both at NLO and N$^2$LO. If the whole expressions for $A_s$ and $\d_{m_s}$ were to be used in the N$^2$LO amplitude, higher order terms would be introduced. Therefore we will differentiate between $A^{NLO}_s$ and $A^{N^2LO}_s$ as well as between $\d^{NLO}_{m_s}$ and $\d^{N^2LO}_{m_s}$, which we will consider as independent parameters. Recall that the 
expression for the phase shift in the low energy region shares the same $A_s$ as in the high energy one, but has an independent $\d_{m_s}$, which we will label $\d^{\slashed{\p}}_{m_s}$.
Because of this shared parameter ($A_s$) we have made a common fit of the low
and high energy region phase shift at each order. The 
low energy region phase shift (calculated in $\slashed{\p}$NNEFT) has been fitted to data in the 0-3MeV range and the high energy region phase shift (calculated in pNNEFT) to data in the 3-50MeV. Results for the $^1S_0$ channel parameters are summarized in Table \ref{taula1}. An alternative fitting procedure was presented in \cite{Tarrus:2009zt}.

The phase shifts are plotted in fig.\ref{1s0nlo}(NLO) and fig.\ref{1s0nnlo}(N$^2$LO) versus center of mass (CM) energy. The low energy region and the high energy region phase shifts have been plotted in the 0-4MeV and 1-50MeV range respectively. 

\begin{table}
\footnotesize{\centerline{
\begin{tabular}{|c|c|c|c|c|c|c|} \hline
     & $A^{NLO}_s$(MeV$^{-1/2}$) & $A^{N^2LO}_s$(MeV$^{-1/2}$) & $\d^{NLO}_{m_s}$(MeV) & $\d^{N^2LO}_{m_s}$(MeV) & $\d^{NLO,\slashed{\p}}_{m_s}$(MeV) & $\d^{N^2LO,\slashed{\p}}_{m_s}$(MeV)  \\ \hline
 NLO & $0.0291$ & & $-1.40$ & & $-3.90$ & \\ \hline
 N$^2$LO & $0.0361$ & $0.0277$ & $-13.7$ & $-17.4$ & $2.10$ & $-1.89$\\ \hline
\end{tabular}}} 
\caption{\footnotesize{Fit values of the parameters for the $^1S_0$ channel.}}
\label{taula1}
\end{table}

\begin{figure}
\centerline{\rotatebox{0}{\resizebox{10cm}{7.5cm}{\includegraphics{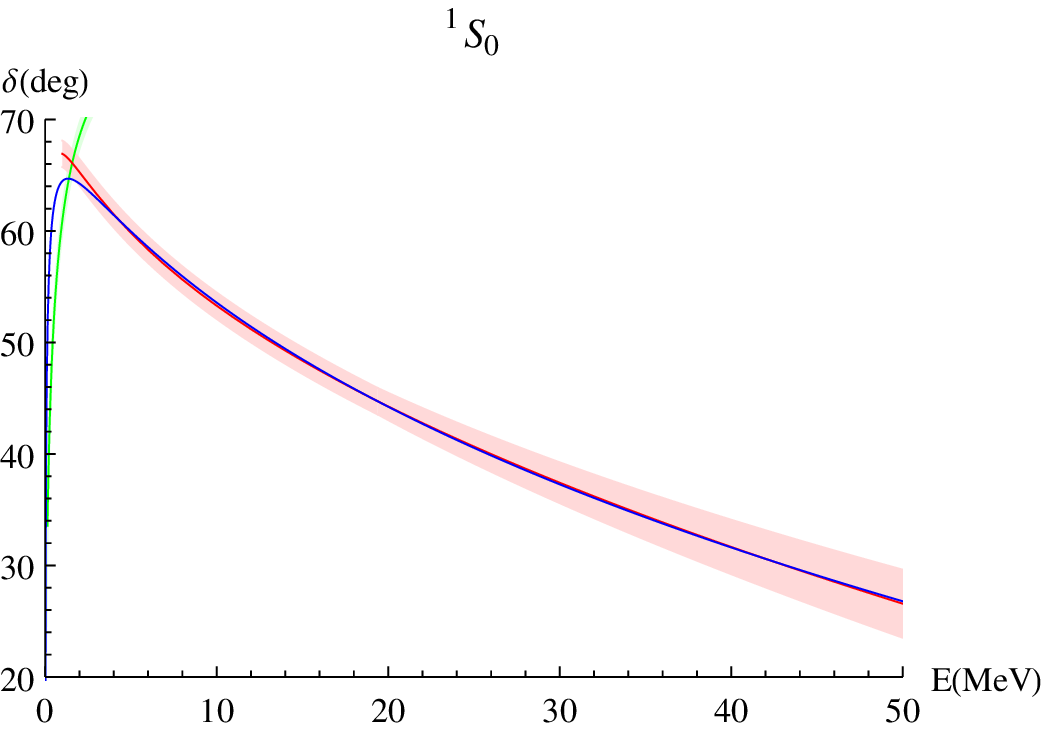}}}}
\caption{'(Color online)'\footnotesize{Plot of the NLO expression for the $^1S_0$  phase shift versus CM energy. The blue line shows the Nijmegen data for the $^1S_0$ phase shift, while the red and green line correspond to the high energy and 
low energy expressions respectively. The fitting procedure is explained in the text. Error bands correspond to $\pm (\frac{m_{\p}}{m_N})^2$ for $p\leq m_{\p}$ and to $\pm (\frac{p}{m_N})^2$ for $p > m_{\p}$ }}
\label{1s0nlo}
\end{figure}

\begin{figure}
\centerline{\rotatebox{0}{\resizebox{10cm}{7.5cm}{\includegraphics{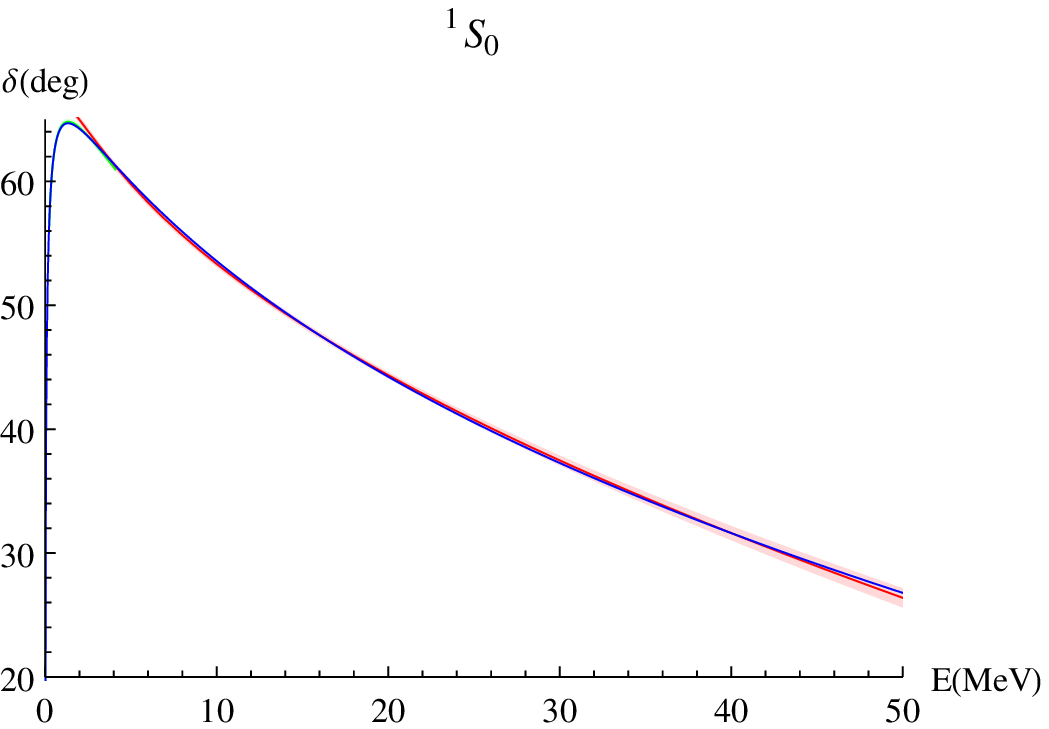}}}}
\caption{'(Color online)'\footnotesize{Plot of the N$^2$LO expression for the $^1S_0$ phase shift versus CM energy. As in the previous figure the blue line shows the Nijmegen data for the $^1S_0$ phase shift. The red line corresponds to the high energy expression and the green one to the 
low energy one (which totally overlaps the data). Error bands correspond to $\pm (\frac{m_{\p}}{m_N})^3$ for $p\leq m_{\p}$ and to $\pm (\frac{p}{m_N})^3$ for $p > m_{\p}$ }}
\label{1s0nnlo}
\end{figure}

\section{The $^3 S_1$-$^3 D_1$ channel}

\paragraph{}In this section we analyze $^3 S_1$-$^3 D_1$ channel. We compare the $^3 S_1$ and $^3 D_1$ phase shifts to data as well as the mixing angle. The usual expression for the S-matrix in this channel,

\be
S=1+i\frac{pm_N}{2\p}\left(
\begin{array}{cc}
\mathcal{A}^{v}	& \mathcal{A}^{mix} \\
\mathcal{A}^{mix} & 	\mathcal{A}^{^3D_1}\\
\end{array}
\right)
=
\left(
\begin{array}{cc}
e^{2i\d^{(^3S_1)}}\cos(2\e)	& ie^{i\d^{(^3S_1)}+i\d^{(^3D_1})}\sin(2\e)\\
ie^{i\d^{(^3S_1)}+i\d^{(^3D_1)}}\sin(2\e) & e^{2i\d^{(^3D_1)}}\cos(2\e)	\\
\end{array}
\right).
\ee

To obtain the phase shift expression at each order we expand both sides in powers $(m_{\p}/\L_{\chi})^n$, as we did in the previous section, and solve the resulting system to obtain $\d^{v,LO}$, $\d^{v,NLO}$ and, $\d^{v,N^2LO}$; $\d^{^3D_1,NLO}$ and, $\d^{^3D_1,N^2LO}$; $\e^{NLO}$ and $\e^{N^2LO}$. There is no $\e^{LO}$ or $\d^{^3D_1,LO}$ due to the fact that $\mathcal{A}^{mix}$ and $\mathcal{A}^{^3D_1}$ start at NLO.

\paragraph{}The fitting procedure for the NLO result is analogous to the one used for the $^1S_0$ channel. For the N$^2$LO one, several changes had to be introduced. A common fit to the
low energy phase shift and to the mixing angle have been made, whereas the high energy phase shift has been left out and fitted independently. This is because all attempts to fit the high energy phase shift together with the other two expressions failed. The N$^2$LO pNNEFT phase shift fit delivers a value for $A^{N^2LO}_v$ (Table \ref{taula3}) which is far away from the expected natural size. We think this is the reason why we were unable to perform a successful common fit: whereas the mixing angle and the 
low energy phase shift favor natural size parameters, the high energy phase shift does not. This is a clear sign that our approach fails to converge in the $^3 S_1$-$^3 D_1$ channel, we will comment on it further in the next section.
Note that the parameter $B'_v/A_v$ only appears in the N$^2$LO the mixing angle. $\e^{NLO}$, $\d^{^3D_1,NLO}$ and, $\d^{^3D_1,N^2LO}$ do not contain free parameters. An alternative fitting procedure was presented in \cite{Tarrus:2009zt}.

The $^3S_1$ phase shifts are plotted in fig.\ref{plotnlo3s1}(NLO) and fig.\ref{plotnnlo3s1}(N$^2$LO), the mixing angle in fig.\ref{mix} and the $^3D_1$ phase shift in fig.\ref{plote3d1}. All $^3S_1$-$^3D_1$ channel plots are versus CM energy. The low energy region and high energy region phase shifts have been plotted in the 0-4MeV and 1-50MeV range respectively, the mixing angle and the $^3D_1$ phase shift have been plotted in the 0-50MeV range. Results for the $^3S_1$-$^3D_1$ channel parameters are summarized in Table \ref{taula2} and Table \ref{taula3}.

\begin{figure}
\centerline{\rotatebox{0}{\resizebox{10cm}{7.5cm}{\includegraphics{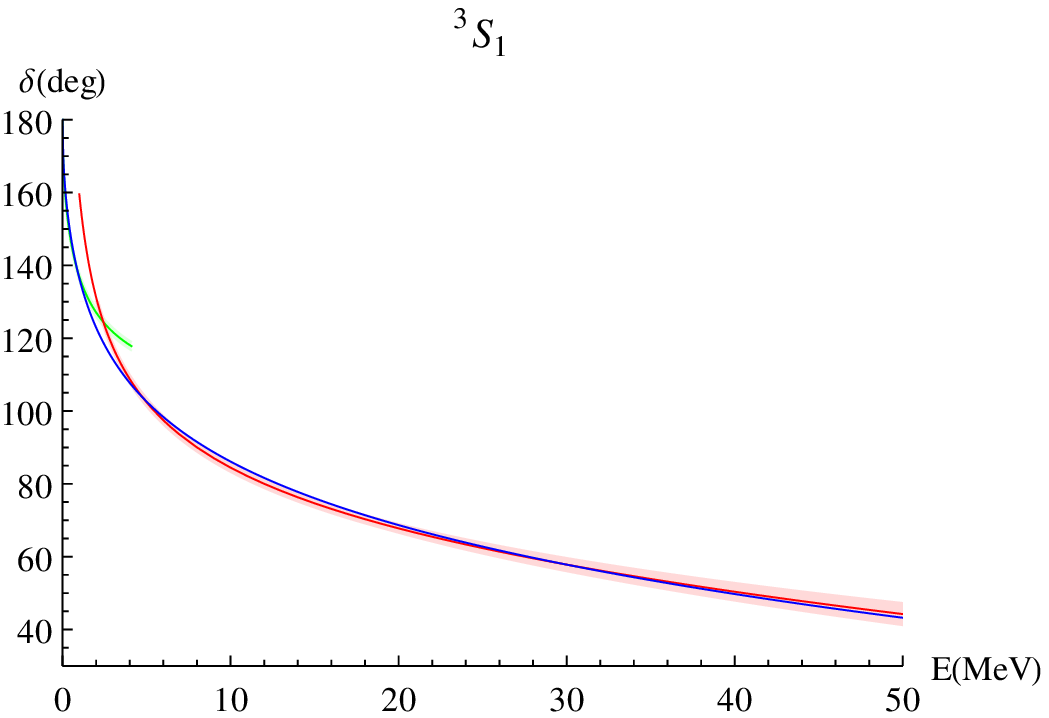}}}}
\caption{'(Color online)'\footnotesize{Plot of the NLO expression for the $^3S_1$ phase shift versus CM energy. The blue line shows the Nijmegen data for the $^3S_1$ phase shift, the red line corresponds to the high energy region expression and the green to the 
low energy region one. The fitting procedure is explained in the text. Error bands correspond to $\pm (\frac{m_{\p}}{m_N})^2$ for $p\leq m_{\p}$ and to $\pm (\frac{p}{m_N})^2$ for $p > m_{\p}$}}
\label{plotnlo3s1}
\end{figure}

\begin{figure}
\centerline{\rotatebox{0}{\resizebox{10cm}{7.5cm}{\includegraphics{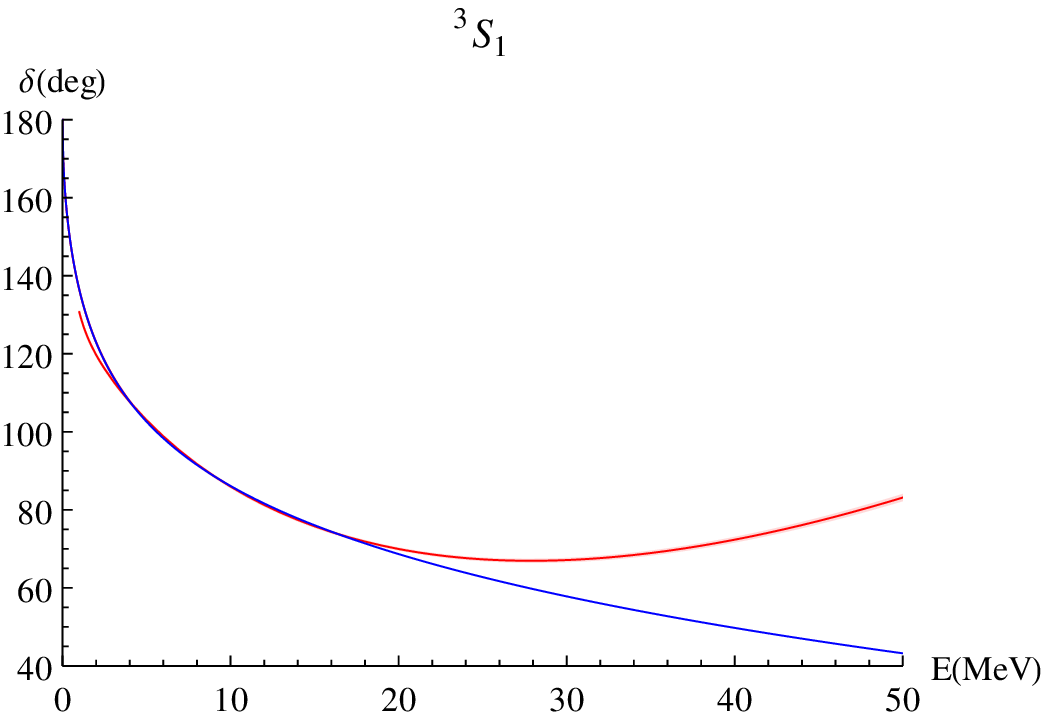}}}}
\caption{'(Color online)'\footnotesize{Plot of the N$^2$LO expression for the $^3S_1$ phase shift versus CM energy. The blue curve is the Nijmegen data for the $^3S_1$ phase shift, while red 
line corresponds to the 
high energy region 
expression.
The curve for the low energy expression totally overlaps with data. The fitting procedure is explained in the text. Error bands correspond to $\pm (\frac{m_{\p}}{m_N})^3$ for $p\leq m_{\p}$ and to $\pm (\frac{p}{m_N})^3$ for $p > m_{\p}$ }}
\label{plotnnlo3s1}
\end{figure}

\begin{figure}
\centerline{\rotatebox{0}{\resizebox{10cm}{7.5cm}{\includegraphics{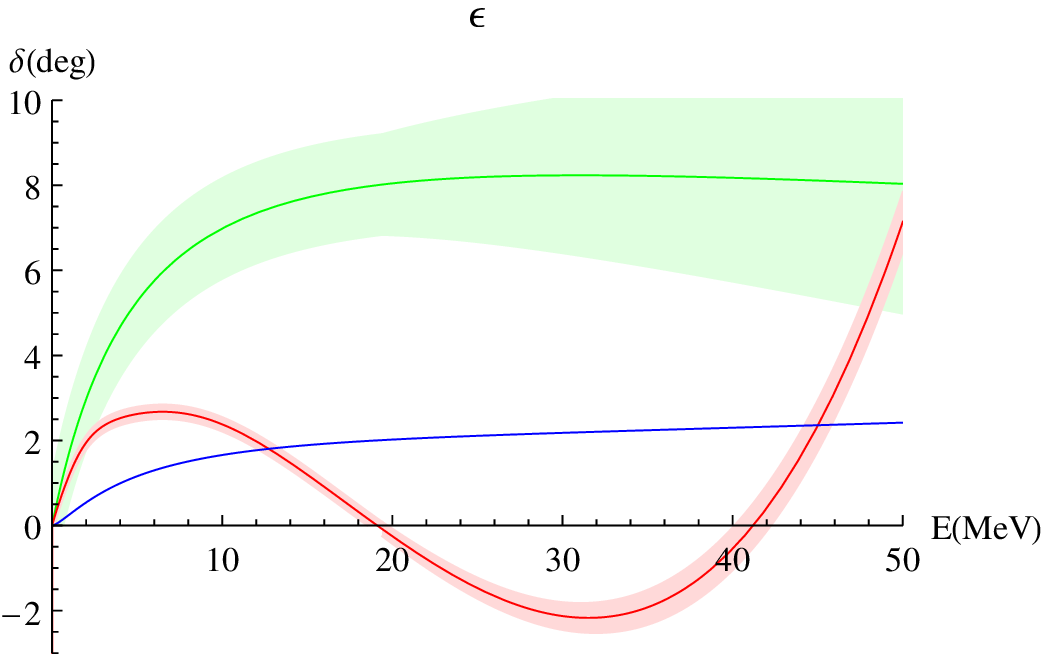}}}}
\caption{'(Color online)'\footnotesize{Plot of the mixing angle versus CM energy. The blue line shows the Nijmegen data, the green and red lines the NLO and N$^2$LO expression respectively. The NLO expression has no free parameters. The free parameters of the N$^2$LO expression have been fitted as explained in the text. The light green (light red) error bands correspond to $\pm (\frac{m_{\p}}{m_N})^{2(3)}$ for $p\leq m_{\p}$ and to $\pm (\frac{p}{m_N})^{2(3)}$ for $p > m_{\p}$.}}
\label{mix}
\end{figure}

\begin{figure}
\centerline{\rotatebox{0}{\resizebox{10cm}{7.5cm}{\includegraphics{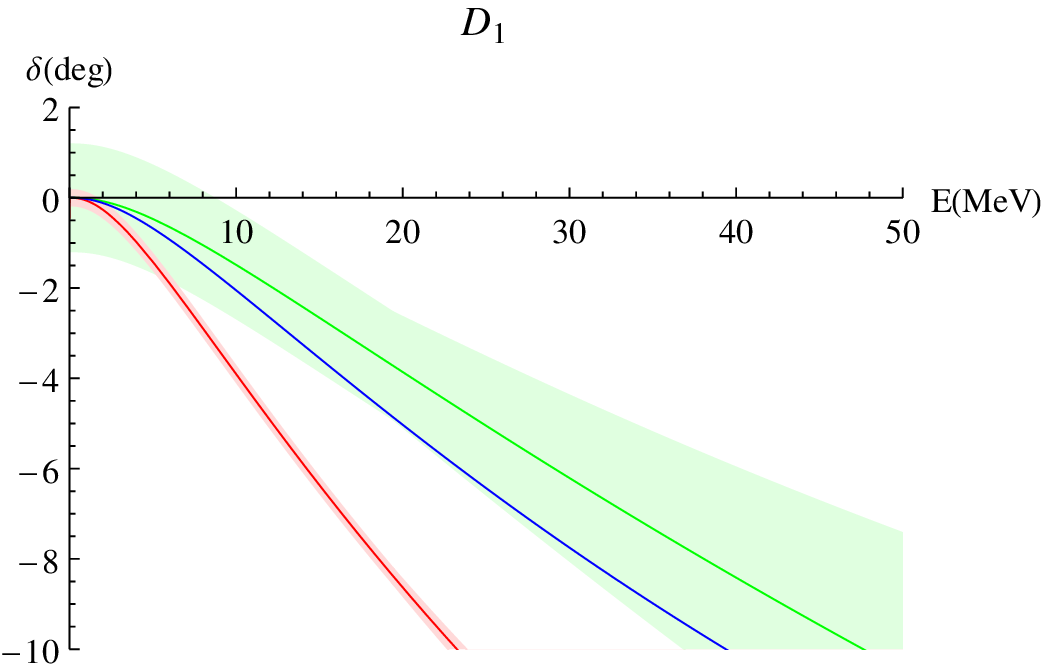}}}}
\caption{'(Color online)'\footnotesize{Plot of the $^3D_1$ phase shift versus CM energy. The blue line shows the Nijmegen data, the green and red lines the NLO and N$^2$LO expression respectively. Neither the NLO or the N$^2$LO expression have free parameters. The light green (light red) error bands correspond to $\pm (\frac{m_{\p}}{m_N})^{2(3)}$ for $p\leq m_{\p}$ and to $\pm (\frac{p}{m_N})^{2(3)}$ for $p > m_{\p}$.}}
\label{plote3d1}
\end{figure}

\begin{table}
\scriptsize{\centerline{
\begin{tabular}{|c|c|c|c|c|c|c|c|} \hline
     & $A^{NLO}_v$(MeV$^{-1/2}$) & $A^{N^2LO}_v$(MeV$^{-1/2}$) & $\d^{NLO}_{m_v}$(MeV) & $\d^{N^2LO}_{m_v}$(MeV) & $\d^{NLO,\slashed{\p}}_{m_v}$(MeV) & $\d^{N^2LO,\slashed{\p}}_{m_v}$(MeV) & $B_v'/A_v$ (MeV$^{-2}$)  \\ \hline
 NLO & $0.0305$ & & $12.14$ & & $8.30$ & & \\ \hline
 N$^2$LO & $0.0431$ & $0.0429$ & $-9.29$ & $-13.3$ & $-23.0$ & $19.9$ & $-1.78\cdot10^{-5}$\\ \hline
\end{tabular}}} 
\caption{\footnotesize{Fit values of the parameters for the $^3S_1-^3D_1$ channel, excluding the N$^2$LO $^3S_1$ phase shift in the high energy region.}}
\label{taula2}
\end{table}

\begin{table}
\footnotesize{\centerline{
\begin{tabular}{|c|c|c|c|} \hline
 $A^{NLO}_v$(MeV$^{-1/2}$) & $A^{N^2LO}_v$(MeV$^{-1/2}$) & $\d^{NLO}_{m_v}$(MeV) & $\d^{N^2LO}_{m_v}$(MeV) \\ \hline
 $0.0206$ & $0.00996$ & $35.3$ & $3.04$ \\ \hline
\end{tabular}}} 
\caption{\footnotesize{Fit values of the parameters delivered by the N$^2$LO $^3S_1$ phase shift in the high energy region}}
\label{taula3}
\end{table}

\section{Conclusions}

\paragraph{}We have calculated the nucleon-nucleon scattering amplitudes for energies smaller than the pion mass in the $^1S_0$ and the $^3S_1-^3D_1$ channels at N$^2$LO in a chiral effective field theory which contains dibaryon fields as fundamental degrees of freedom, the NNEFT. The large scattering lengths in the $^1S_0$ and the $^3S_1$ channels force the dibaryon residual masses to be much smaller than the pion mass. We organize the calculation in a series of effective theories, which are obtained by sequentially integrating out higher energy and momentum scales. We first integrate out energy scales of the order of the pion mass. This leads to an effective theory with dibaryon and nucleon fields, pNNEFT. The latter interact through potentials. For three momenta of the order of the pion mass, the scattering amplitudes are calculated in this effective theory. For three momenta much smaller than the pion mass, it is convenient to further integrate out three momenta of the order of pion mass, which leads to the $\slashed{\p}$NNEFT, and carry out the calculations in the latter. By splitting the calculation in this way we can take advantage of the modern techniques of the threshold expansions and dimensional regularization so that all integrals only depend on a single scale  \cite{Beneke:1997zp,Manohar:1997qy,Pineda:1998kn}. There is no need to introduce a PDS scheme \cite{KSW}. The technical complexity of the N$^2$LO calculation is similar to the one in the KSW scheme \cite{Fleming:1999ee}, but our final expressions are simpler. 

The numerical results for the phase shifts and mixing angle are also similar to the ones obtained in the KSW approach. Hence a good description of the $^1S_0$ channel is obtained 
, but for the $^3S_1-^3D_1$ channel our results also fail to describe data.
The $^3S_1$ phase shift shows a good agreement with data up to center of mass energies of $50 MeV$ at NLO, but at N$^2$LO the range of the agreement is reduced, up to $20 MeV.$ only, even when the high energy region of this channel is fitted independently, as in the plot of fig.\ref{plotnnlo3s1}. The mixing angle poorly agrees with data, but shows a marginal improvement from NLO to N$^2$LO. The N$^2$LO mixing angle plot is significantly different from the one shown in \cite{Fleming:1999ee}, this is a consequence of making a common fit of the $^3S_1$ phase shift in the low energy region and the mixing angle. For a different fitting approach with plot closer to \cite{Fleming:1999ee} see \cite{Tarrus:2009zt}. Comparison with data for the $^3D_1$ phase shift it is never good. Particularly worrying is the fact that for the $^3S_1$ and the $^3D_1$ phase shift the N$^2$LO calculation compares worse to data than the NLO one.  The reasons of this failure can be traced back to the iteration of the OPE potential , the first diagram in Fig.\ref{nnlopp}, which gives a very large contribution \cite{Fleming:1999ee}. This may be interpreted as an indication that pion exchanges must be iterated at all orders, as originally proposed by Weinberg \cite{Weinberg}. However, the removal of the cut-off in this approach appears to require an infinite number of counterterms, one for each partial wave \cite{PavonValderrama:2005gu,Nogga:2005hy,PavonValderrama:2005uj} (see also \cite{Beane:2001bc}). A very recent proposal, which keeps the essentials of KSW counting, consist in introducing a Pauli-Villars regularization for the pion exchanges and staying at the regularized level \cite{Beane:2008bt}. This seems to produce slightly better results, but it is unclear at the moment that, staying at regularized level, this approach is superior to Weinberg's one \cite{Epelbaum:2008ga} (see \cite{Yang:2009kx,Yang:2009pn} for very recent efforts on the renormalization of Weinberg's approach).

\paragraph{}Before closing we would like to add a few remarks to the current situation. Starting at N$^2$LO the expansion parameter in our approach is $\sqrt{p/\Lambda_\chi}$ rather than  $p/\Lambda_\chi$  (in fact it is an accident due to Wigner symmetry that up to N$^2$LO the expansion parameter is the latter). The N$^3$LO calculation, i.e. ${\cal O}(m_\pi^{3/2}/\Lambda^{7/2})$ would be relatively simple in our approach. Since fractional powers only arise from diagrams involving radiation pions, it would basically consist of taking into account $E/m_\pi$ corrections to the diagrams in Fig.\ref{radpion}.a, and considering a further potential pion exchange or and (extra) internal energy insertion (a cross) in the diagrams in Fig.\ref{radpion}.b and Fig.\ref{radpot}. The outcome of the calculation, however, would amount to redefinitions of previously existing parameters, and hence it would not produce an improvement in the description of data. Qualitative changes are expected at N$^4$LO. It is at this order, for instance, that the two pion exchange potential first enters the calculation. The N$^4$LO calculation appears feasible in our formalism, but would require a major effort.


\section*{Acknowledgments}
\indent
We thank Enrique Ruiz Arriola for explanations on refs. \cite{PavonValderrama:2005gu,PavonValderrama:2005uj}. We have benefited from the participation in the ECT* workshop "Bound States and Resonances in Effective Field Theories. We acknowledge financial support from the CIRIT grant 2005SGR00564 (Catalonia), MICINN grants FPA2007-60275/, FPA2007-66665-C02-01/ (Spain), the RTN Flavianet MRTN-CT-2006-035482 and the European Community-Research Infrastructure
Integrating Activity
"Study of Strongly Interacting Matter" (acronym HadronPhysics2, Grant Agreement
n. 227431)
 (EU). JT has been supported by a FI grant from Departament d'Universitats, Recerca i Societat de la Informaci\'o of the Generalitat de Catalunya and a MICINN FPU grant ref.AP2007-01002.

\appendix
\renewcommand{\thesection}{\Alph{section}.}
\renewcommand{\theequation}{\thesection \arabic{equation}}
\section{Matching prescription}

In the matching calculations between NNEFT and pNNEFT there are regions in the integrals in which $k^0\sim {\bf k}\sim m_\pi$. In these regions the kinetic term of the nucleons is parametrically smaller than the energy and hence, following the ideas of the threshold expansions \cite{Beneke:1997zp}, it must be expanded. However, when one does so in the two nucleon sector one often encounters pinch singularities. It was argued in \cite{Pineda:1998kn} that the pinch singularities can be rearranged in such a form that they exactly match the pinch singularities of the effective theory when the kinetic term is also expanded in the latter. Consequently, once this rearrangement is done, pinch singularities can be safely ignored. As an example, let us consider the one loop contribution to the nucleon-dibaryon vertices produced by a pion exchange. External three momenta are of the order of $m_\pi$ and external energies much smaller than it. Once small scales are expanded we are faced with the following integral,
\be
\int {d^Dk\over (2\pi )^D }{k^ik^j \over k^2-m_\pi^2+i\epsilon}{1\over k^0+i\epsilon} {1\over -k^0+i\epsilon}       
=
-\int {d^Dk\over (2\pi )^D} {k^ik^j \over {\bf k}^2+m_\pi^2}{1\over k^0+i\epsilon} {1\over -k^0+i\epsilon}
\ee
The last expression matches exactly the contribution of the OPE potential to the vertex in pNNEFT, if the kinetic terms of the nucleons are correspondingly expanded. Hence, in this case there is no contribution to the matching.

Note also that with this prescription  the size of each diagram in the $k^0\sim {\bf k}\sim m_\pi$ region is easily estimate since the integrals depend on a single scale. For instance, the first diagram in Fig. 3 (a) has also a contribution in this region but it can be easily seen to be higher order. This is not so if the on-shell prescription is used \cite{Fleming:1999ee}.

\section{The complete NLO Lagrangian in the $N_B=2$ sector}

We list here all the operators of the NLO Lagrangian in the $N_B=2$, many of which do not contribute to our calculations. We use for organization purposes the standard chiral counting, namely $\partial_0 \sim \partial_i = {\cal O}(p)$ and the quark mass matrix ${\cal M}= {\cal O}(p^2)$

\subsection{The dibaryon Lagrangian}

The full list of operators in ${\cal L}_{{\cal O}(p^2)}$ of (\ref{ordrep2}) follows
\bea
&& Tr[D_s(u\mathcal{M}^{\dag}u+u^{\dag}\mathcal{M}u^{\dag})D^{\dag}_s] \quad ,\quad Tr[D^{\dag}_s(u\mathcal{M}^{\dag}u+u^{\dag}\mathcal{M}u^{\dag})D_s]\quad ,\quad \nn\\
&& Tr[D^{\dag}_sD_su_0u_0] \; , \;
Tr[D^{\dag}_sD_su_iu_i] \; ,\;
Tr[D_sD^{\dag}_su_iu_i] \; ,\;
Tr[D_s^{\dag}u_0D_su_0]\; ,\; Tr[D_s^{\dag}u_iD_su_i]\nn \\
&&\vec{D}^{\dag}_v\cdot\vec{D}_vTr[u^{\dag}\mathcal{M}u^{\dag}+u\mathcal{M}^{\dag}u] \quad ,\quad
\vec{D}^{\dag}_v\cdot\vec{D}_vTr[u_0u_0]\quad ,\quad
\vec{D}^{\dag}_v\cdot\vec{D}_vTr[u_iu_i]\nn\\
&& (D_v^{i\dag}D_v^{j}+D_v^iD_v^{j\dag})Tr[u_iu_j]\quad ,\quad Tr[D_s^{\dag}{\vec u}\times {\vec u}]\vec{D}_v + {\rm h.c.} \\
&& {\vec \pa}\vec{D}^{\dag}_v Tr[u_0D_s]+ {\rm h.c.} \quad ,\quad \vec{D}^{\dag}_v Tr[u_0{\vec d}D_s]+ {\rm h.c.}\nn\\
&& Tr[{\vec d}D^{\dag}_s{\vec d}D_s] \quad ,\quad \left( {\vec \pa}\vec{D}^{\dag}_v\right)\left({\vec \pa}\vec{D}_v\right) \quad ,\quad {\vec D}^{\dag}_v {\vec \pa}^2 \vec{D}_v\nn
\eea
In Ref. \cite{Soto:2007pg} terms mixing the scalar and vector dibaryon as well as terms with space derivatives on the dibaryon field were not displayed\footnote{The two additional terms that were displayed in \cite{Soto:2007pg} turn out to be redundant.}.

\subsection{The dibaryon-nucleon vertex}

The full list of operators in $\mathcal{L}_{DN}^{(2)}$ of (\ref{dn2}) follows (hermitian conjugates are omited)

\bea
&& (N^{\dag}\s^2\t^a\t^2{\vec D}^2N^*)D_{s,a} 
 \quad ,\quad
(N^{\dag}\t^2\vec{\s}\s^2{\vec D}^2N^*)\cdot\vec{D}_v \quad ,\quad
 (D_i N^{\dag}\t^2\s^i\s^2D_j N^*)D^j_v \nn\\
&& (N^{\dag}\s^2D_s\t^2N^*)\left( Tr(u_0u_0) \quad ,\quad  Tr(u_iu_i) \quad ,\quad Tr(u\mathcal{M}^{\dag}u+u^{\dag}\mathcal{M}u^{\dag}) \right) \\
&& N^{\dag}\left( u_0D_su_0 \; , \; u_iD_su_i\; , \; D_su\mathcal{M}^{\dag}u \; , \; D_s u^{\dag}\mathcal{M}u^{\dag} \; , \;  u^{\dag}\mathcal{M}u^{\dag} D_s\; , \; u\mathcal{M}^{\dag}u D_s \right) \t^2\s^2 N^\ast\nn\\
&& N^{\dag}\s^i\left( \d^{ij}u\mathcal{M}^{\dag}u \; , \; \d^{ij}u^{\dag}\mathcal{M}u^{\dag} \; , \;  \epsilon^{ijk}u_k\; , \; u_iu_j \; , \; \epsilon^{ijk}D^k u_0\; , \; \epsilon^{ijk}u_0D_k\right) \t^2\s^2 N^\ast D_v^j \nn\\
&&N^{\dag}\t^2\s^i\s^2N^*\left( \d^{ij}Tr(u_0u_0) \; , \,  \d^{ij}Tr(u_ku_k) \; , \, \d^{ij}Tr(u\mathcal{M}^{\dag}u+u^{\dag}\mathcal{M}u^{\dag}) \; , \,  Tr(u_iu_j)\right) {D}_v^j\nn\\
 && N^{\dag}\s^i\left( u^iD_s\quad ,\quad D_s u^i \quad ,\quad  \epsilon^{ijk}u_ju_kD_s\quad ,\quad \epsilon^{ijk}u_jD_su_k \quad ,\quad \epsilon^{ijk}D_su_ju_k \right) \t^2\s^2 N^\ast\nn\\
&& N^{\dag}\left( u^i\quad ,\quad   \epsilon^{ijk}u_ju_k\quad ,\quad u_0 D^i \quad ,\quad D^i u_0 \right) \t^2\s^2 N^\ast D_v^i\nn
 \eea


\end{document}